\documentclass[%
 preprint,
amsmath,amssymb,
aps,
pra,  
longbiblioography,
]{revtex4-2}

\usepackage{graphicx}% Include figure files
\usepackage{dcolumn}% Align table columns on decimal point
\usepackage{bm}% bold math
\usepackage{longtable}
\usepackage{verbatim}
\usepackage{setspace}

\usepackage{xcolor}	% Colored text

\newcommand{\eEDM}{{\em e}EDM}

\newcommand{\wn}{cm$^{-1}$}
\newcommand{\PbF}{$^{207}$PbF}
\newcommand{\myremm}[1]{\textcolor{blue}{#1}}
\newcommand{\ecm}{\ensuremath{e {\cdotp} {\rm cm}}}
\newcommand{\de}{d_\mathrm{e}}

\begin{document}

\preprint{APS/123-QED}

\title{Rotational and Near-IR Spectra of PbF: Characterization of the Coupled $X_1\,^2\Pi_{1/2}$ and $X_2\,^2\Pi_{3/2}$ States}% Force line breaks with \\

\author{Sean Jackson} \thanks{Graduated Pomona College, 2023}
\author{Luke Kim}\thanks{Graduated Pomona College, 2018}
\author{Andreas Biekert}\thanks{Graduated Pomona College, 2016}
\author{Alex Nguyen}\thanks{Graduated Pomona College, 2018}
\author{Richard J Mawhorter}
 \email{rjm04747@pomona.edu}
\affiliation{Department of Physics and Astronomy, Pomona College, Claremont, CA 91711, USA}%Lines break automatically or can be forced with \\
\author{Trevor J. Sears}%
 \email{trevor.sears@stonybrook.edu}
\affiliation{Chemistry Department, Stony Brook University, Stony Brook, NY 11794-3400}%
\author{Leonid V. Skripnikov}
\author{Vera V. Baturo}
\author{Alexander N. Petrov}
 \email{alexandernp@gmail.com}
 \affiliation{Federal State Budgetary Institution ``Petersburg Nuclear Physics  Institute'', Gatchina, Leningrad district 188300, Russia}
\affiliation{Saint Petersburg State University, St. Petersburg, 199034,
Russia}

 %This line break forced% with \\}%
\author{Jens-Uwe Grabow}
 \email{jens-uwe.grabow@pci.uni-hannover.de}
\affiliation{
 Gottfried Wilhelm Leibniz Universit\"at, Institut f\"ur Physikalische Chemie and Elektrochemie, Hannover, 30167, Germany\\ }

\date{\today}% It is always \today, today,

\begin{abstract}
Observations of the rotational spectrum of lead monofluoride, PbF, have been extended up to transitions in the \textit{v} = 7 level for $^{208}$PbF in the lowest $X_1\,^2\Pi_{1/2}$ state of the radical and \textit{v} = 5 for the $^{207}$Pb and $^{206}$Pb isotopologues.  The data also include a few measurements for $^{204}$PbF in \textit{v} = 0. These new measurements have been combined with existing near-IR measurements of the $X_2 - X_1$ fine-structure transition and a simultaneous multi-isotope fit of the data to an effective isotope-independent ro-vibronic Hamiltonian has been carried out. The resulting parameters fully characterize the vibrational, rotational and hyperfine structure of the combined $X_1 \, / \, X_2$ state of the radical.  A pair of opposite parity levels with total angular momentum quantum number, $F=1/2$, in the lowest rotational level, $J=1/2$ of \PbF \,are close in energy and their spacing decreases with vibrational excitation. The experimental results show the spacing decreases to less than 20 MHz at $v=7$ and 8. The experimental work is complemented by new \textit{ab initio} calculations which support the results and allow predictions outside the experimental data range. The calculated radiative lifetimes of the relevant vibrationally excited states are of the order of 50 ms. This work was motivated by interest in using \PbF\, as a vehicle for future probes of the standard model of physics such as placing limits on the electron's electric dipole moment (\eEDM), molecular charge-parity non-conservation and Born-Oppenheimer breakdown effects for example.  

\end{abstract}

\maketitle

%\tableofcontents

\section{\label{sec:intro1}Introduction}

In 1978 the investigation of the electron's dipole moment (\eEDM) and other parity non-conserving (PNC) effects in lead monfluoride, PbF, was suggested by Labzowsky and Gorshkow  \cite{Labzowsky:78,Gorshkov:79} and Sushkov and Flambaum \cite{Sushkov:78}
 who showed how these effects might be enhanced in diatomic radicals like BiS and PbF due to the closeness of levels of opposite parity in $\Omega$-doublets having a $^2\Pi_{1/2}$ ground state. The first two-step {\sl ab~initio}
 calculation of PNC effects in PbF initiated by Labzowsky was completed by Titov et al. \cite{Kozlov:87}.Much later, PbF emerged as a potential candidate molecule for experimental and theoretical studies of physics beyond the standard model when Shafer-Ray et al. \cite{Shafer-Ray:06} predicted that the electric field-dependent $g-$factor could cross zero at high electric fields. \\
 
This result suggested that the molecule could be an attractive target for experiments designed to place a limit on the size of the \eEDM.  In order to design such an experiment, knowledge of the detailed internal energy level structure of the molecule is necessary.  In 2011, Mawhorter et al. \cite{Mawhorter2011} reported measurements of rotational transitions among the lowest rotational levels of the molecule that more fully characterized the ground state fine and hyperfine split energy level structure for all stable PbF isotopologues in natural abundance, $^{208}$PbF (52.4 percent), $^{207}$PbF (22.1 percent),$^{206}$PbF (24.1 percent), and $^{204}$PbF (1.4 percent).  As just noted, due to the intrinsic orbital angular momentum in a $^2\Pi$ state, molecules containing even lead  isotopes exhibit distinct $\Omega$-doubling between opposite parity levels in the ground $X_1\, ^2\Pi_{1/2}$ state with small $^{19}$F nuclear hyperfine splittings. The handedness of this orbital angular momentum relative to the molecular axis provides inherent symmetry sensitivity akin to the handedness arising from the bending mode degeneracy in molecules like YbOH \cite{Kozyryev2017, Petrov:2022, Jadbabaie2023, Petrov:2024}, but without the attendant polyatomic complications.  For the even PbF isotopologues, opposite parity fine and hyperfine split rotational levels are well separated.  In \PbF\, though, the large $^{207}$Pb ($I_{Pb} = \frac{1}{2}$) hyperfine interaction causes a near cancellation of the $\Omega$-doubling splitting in some levels resulting in a pair of levels of opposite parity only 266 MHz apart in the lowest rotational and vibrational level.  Alphei et al. \cite{Lukas2011} showed, on the basis of these results, how the  $^{207}$Pb-containing isotopologue was a particularly strong candidate for experiments designed to probe certain symmetry-breaking effects. Follow-up studies included a determination of centrifugal distortion parameters for the hyperfine structure constants \cite{Petrov2013} needed to refine hyperfine splitting calculations, and an investigation of the theoretical and experimental  $^{208}$PbF g-factors using Zeeman data \cite{Skripnikov2015}.  \\

 In 2023 the JILA group obtained a new constraint on the  \eEDM, $|\de|<4.1\times 10^{-30}$ \ecm\ (90\% confidence) \cite{newlimit1}, using $^{180}$Hf$^{19}$F$^+$ ions trapped by a rotating electric field. This result can be compared with the latest ACME collaboration constraint obtained in 2018, $|d_e| < 1.1\cdot 10^{-29}\ e\cdot\textrm{cm}$ in a ThO beam experiment \cite{ACME:18}.  In both $^{180}$Hf$^{19}$F$^+$ and ThO the measurements were performed on the metastable excited electronic $^3\Delta_1$ states. The great progress in the \eEDM~ search using these molecules is closely related to their $\Omega$-doubling structure. As was shown in Ref. \cite{DeMille2001} the $\Omega$-doublet structure is arranged in such a way that $e$EDM contribution to the splitting is opposite to many systematic effects including the one related to stray magnetic fields. Thus, subtracting the measured energy splittings in the two $\Omega$-doublet states suppresses many systematic effects in the experiment. Baturo et al. \cite{Baturo2021} showed the existence of a similar energy level structure for \PbF. \, Equally important, the small zero-field splitting between the opposite parity pair of levels in the lowest rotational level of \PbF \, actually decreases with increasing vibrational excitation, \cite{Mawhorter2018} reaching a minimum between \textit{v} = 7 and 8 where the levels in question cross in energy.  \\

The smaller the zero-field splitting between the opposite parity levels the smaller electric field required to polarize the molecule which reduces potential measurement systematics. Baturo et al. \cite{Baturo2021} also showed that while the $g-$factor in the lowest state of the molecule does not vanish, it is still very small thus reducing potential systematic errors in any future \eEDM\, measurement due to stray magnetic fields. One can note that the set up of the experiment on trapped  HfF${^+}$ ions is very different from molecular beam experiments on ThO and (potentially) PbF. However, it is important to use different molecular systems to confirm the reliability of the measured \eEDM ~limits. The main advantage of $^{207}$PbF as compared to ThO is that the \eEDM~ sensitive state of \PbF\, is in its ground electronic state. \myremm{Even the excited vibrational levels, where the parity spacing is minimized, have lifetimes that are far longer (see Table \ref{tab:VibEngLife} below) than the excited state lifetime of ThO ($\sim 1 $ msec).  Level lifetimes directly translate to coherence times which linearly affect potential experimental sensitivity.} \\

\PbF\, in its lowest $X_1$ state is also a good candidate for experiments designed to search for temporal variation in fundamental constants \cite{Flambaum2013} as well as molecular level consequences of an anapole moment of the $^{207}$Pb nucleus \cite{Lukas2011}. Recent work by Baturo et al. \cite{Baturo2021} has summarized the relevant spectroscopic work up to 2022.  Since 2022, recent work by Luan et al. \cite{Luan_2024} investigated the suitability for laser cooling of the $X_2-X_1$ near-IR electronic transition that is a focus of the present work.  They showed that the highly diagonal nature of the band system made it possible to propose a four-laser scheme to reach submicro-Kelvin temperatures. Zhu et al. \cite{Zhu2022} recorded and analyzed a region of the $B\,^2\Sigma^+ - X_1\,^2\Pi_{1/2}$ band system in a jet-cooled sample of PbF. The analysis combined published near-IR data \cite{Ziebarth1998} and microwave data \cite{Mawhorter2011} together with their new measurements, but was much less comprehensive and does not include the new data in the present work which also combines data for all the naturally occurring isotopes of the molecule. \\

Here, we report the measurement of many new hyperfine-split rotational transitions in all the naturally-occurring isotopic modifications of PbF. These new data include information on excited vibrational levels within the $X_1$ state that permit the experimental determination of the variation in energy of the parity and hyperfine-split levels in the lowest rotational level of \PbF\, with vibrational excitation.  The data have been combined with existing and some newly assigned near-IR spectral data for the molecule, and a combined analysis determines the parameters appearing in the isotope-independent effective Hamiltonian describing the molecule.  In the form of the Hamiltonian model used, the energies resulting from a Dunham-type expansion of parameters $Y_{lm}$, described in detail below, are fitted to the measured transitions. The spectroscopic analysis identified several Born-Oppenheimer breakdown contributions to the molecular energy levels.  These include a non-zero isotopic variation in the equilibrium spin-orbit splitting, and field shifts due to finite-sized nucleus effects which cause systematic isotopic variation beyond the expected isotopic mass relationship in the rotational constant, among other parameters.  These shifts have also been investigated theoretically. The results will be useful for predicting the energies of levels outside the set included in the present data, as well as the energies of other isotopologues, in particular $^{205}$PbF where the odd lead nucleus with spin $I= \frac{5}{2}$  has a quadrupole moment and a half life of 17.3 million years. \\

The present work establishes the size of the splitting between the near-degenerate parity pair of \PbF\,levels, and its vibrational level dependence, to a precision and accuracy of a few kHz, as well as the energies of the rotational-vibration levels in the combined $X_1 \, / \, X_2$ fine-structure pair of levels in the ground $^2\Pi$ state of the molecule up to rotational quantum number, \textit{N}, $\approx$ 60 and vibrational levels, \textit{v} $\le$ 8.  The splitting between the near-degenerate parity pair with total angular momentum quantum number, \textit{F},\, =\,1/2 in \PbF\, reaches a minimum of 15 MHz at \textit{v} = 8. The experimental measurements are complemented by new \textit{ab initio} calculations that compare well to the experimental results and permit extrapolation of the experimental data to still higher energy levels. In the spirit of the earlier work on the $g-$factor variation with electric field \cite{Baturo2021}, we have investigated the variation in the higher vibrational levels where the parity spacing is at its minimum.  We also calculate the radiative lifetime for the \textit{v}\,=\,8 level of $X_1$ \PbF \,in question to be 38 ms, sufficient for future precision measurements that are not restricted by lifetime-limited experimental coherence times.  \\

\section{\label{sec:Expt}Experimental Details}
We have measured approximately 130 hyperfine-split rotational transitions among the lowest few rotational levels of the PbF radical in its ground, $X_1 \, ^2\Pi_{1/2}$ state. The measured frequencies are given in the Appendix and together with the near-IR data included in the analysis, see section \ref{sec:Anal}, are also available in a Pomona College thesis \cite{Jacksonthesis}. The data include transitions at frequencies between approximately 4 and 26 GHz covering the operating range of the spectrometer.  While these are primarily between the lowest J = {3/2} and {1/2} levels, the large $\Omega$-doubling intervals of 4, 8, 12, and 16 GHz with increasing J enable access to the J = {5/2} and {7/2} manifolds as well. The structure of the lowest few energy levels in the molecule is illustrated in Figure 1 of reference \cite{Mawhorter2011}, which shows the profound influence of the spin {1/2} $^{207}$Pb nucleus.  Here the near-degenerate $^{207}$PbF parity pair are the levels labelled 3 and 4.  This reference also includes a description of the form and effective parameters in the hyperfine Hamiltonian required to model the energy level structure. \\

The Fourier transform microwave (FTMW) spectrometer used has been described previously \cite{GrabowCOBRA1996, GrabowCOBRA2005} and PbF was synthesized \textit{in situ} by laser vaporization of a sample of natural abundance Pb metal in the presence of a dilute sample of SF$_6$ entrained in neon gas prior to a supersonic free-jet. Frequencies were determined by the Fourier transform of the free induction decay from the resonator antenna.  This enables sub-kHz resolution for transitions with good signal to noise.  \\

\begin{figure*}[t!]
%    \centering
%    \begin{subfigure}[t]{0.4\textwidth}
%        \centering
        \includegraphics[height=5.5cm]{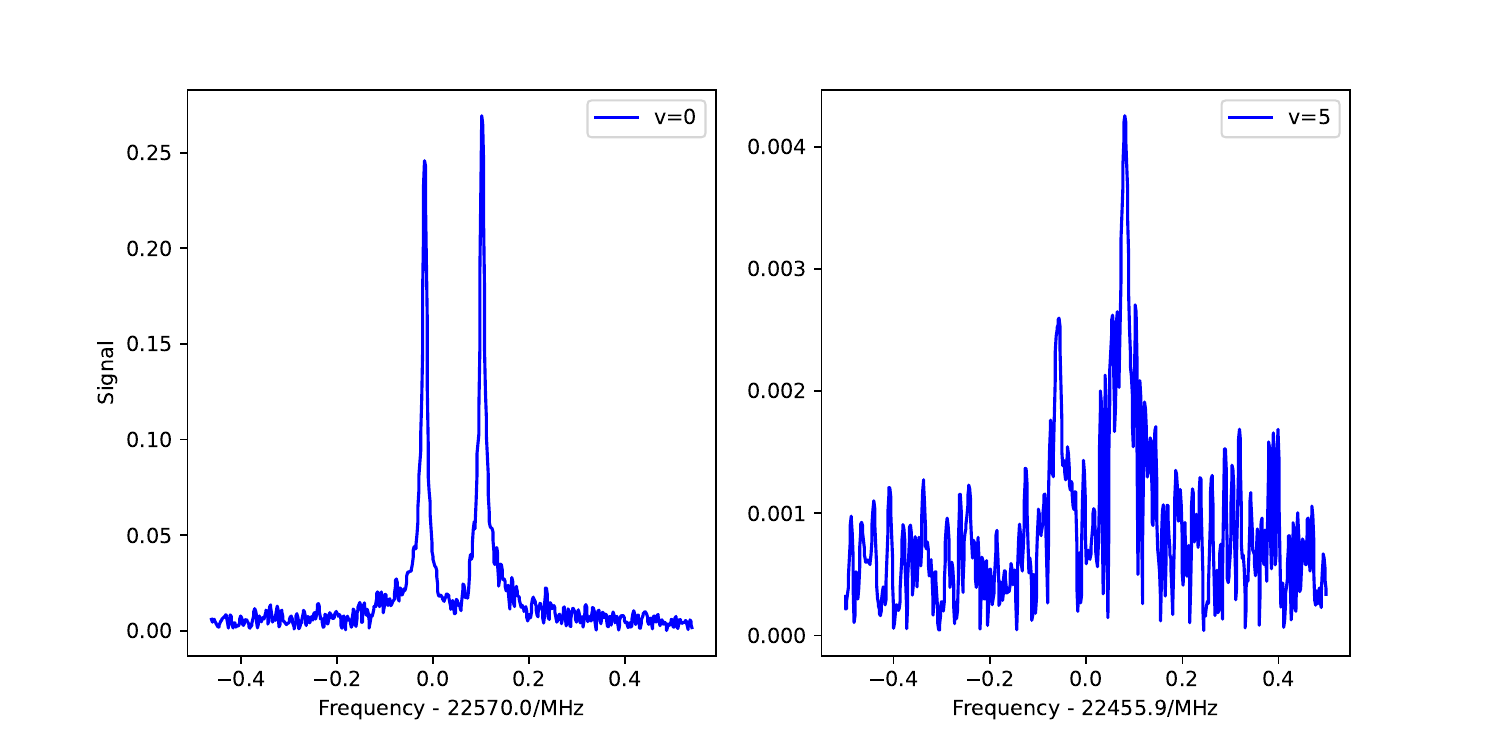}
    \caption{\label{fig:FTMWData}\footnotesize{Examples of FTMW transitions in \textit{v} = 0 and 5 vibrational levels of $^{207}$PbF. The resonance signal appears twice due to the Doppler shift of the expanding molecular jet's emission with respect to the resonator mode's propagation}. The rest frequency is the arithmetic mean of the two frequencies. The experimental conditions were optimized for each transition.}
\end{figure*}

    Importantly, we have been able to improve the spectrometer's capability to record transitions in excited vibrational levels that, even after creation of a laser plasma, suffer from low population after collisional transfer during supersonic expansion. This has been compensated for by a direct current  discharge\cite{Bizzocchi2007} boosting the internal energy of the molecular sample prior to the expansion. This enabled a larger vibrational data set.  \myremm{There are potentially 20 J = {3/2} $\rightarrow$ {1/2} observable transitions among the three most abundant PbF isotopologues.  In vibrationally-excited levels 20, 15, 12, 9, and 6 were measured in the levels \textit{v} = 1-5, respectively. Transitions in \textit{v} = 6 and \textit{v} = 7 were detected in the most abundant $^{208}$PbF and a few transitions in the low-abundance variant, $^{204}$PbF were also detected.}  The different mass isotopologues have predictably different transition frequencies and spectral assignments were usually straightforward.  The requirement that all the different isotopic species data are consistent in the overall fitting provides strong limits against possible mis-assignments.  Examples of the spectra involving \textit{v}\,=\,0 ground state and \textit{v}\,=\,5 excited vibrational levels of \PbF\, are shown in Figure \ref{fig:FTMWData}.\\

\section{\label{sec:Theory}Theoretical Methods}
To calculate potential energy curves and the dependence of the molecule-frame dipole moment on the internuclear distance, we employed the relativistic two-component coupled cluster method with single, double, and perturbative triple excitation amplitudes, known as CCSD(T)~\cite{Bartlett:2007,Visscher:96a}. In these correlation calculations, we excluded 60 inner-core electrons using the valence part of the relativistic effective core potential approach~\cite{Titov:99,Mosyagin:10a,Mosyagin:16}. The basis set for Pb was constructed following the procedure outlined in Ref.~\cite{Skripnikov:13a} and consisted of 22 s$-$, 23 p$-$, 16 d$-$, 8 f$-$, 4 g$-$, and 3 h$-$type Gaussian functions. For F, we utilized Dyall's uncontracted augmented all-electron quadruple-zeta basis set AAE4Z~\cite{Dyall:2016}.  The results were also used to calculate lifetimes of excited vibrational states. To estimate the uncertainty of the lifetimes we compared results obtained at the CCSD(T) and at the coupled cluster singles and doubles, CCSD~\cite{Bartlett:2007,Visscher:96a} levels. \\

To calculate the dependence of the hyperfine structure parameters on the internuclear distance, we employed the 4-component relativistic CCSD(T)~\cite{Visscher:96a} approach within the Dirac-Coulomb Hamiltonian. For Pb, we utilized Dyall's uncontracted AAE3Z~\cite{Dyall:2016} basis set, while for F, the AE3Z~\cite{Dyall:2016} basis set was used. In the correlation calculation, we excluded the $1s..3d$ electrons of Pb. 
The energy cutoff for virtual orbitals was set to 300~Hartree.
Relativistic electronic structure calculations were performed using the {\sc dirac}~\cite{DIRAC19,Saue:2020} and {\sc mrcc}~\cite{MRCC2020,Kallay:1,Kallay:2} codes. The eigenvalues and eigenfunctions of the lead monofluoride
molecule required for calculation of the sensitivities to variation of fundamental constants and to the \eEDM\  were obtained by numerical diagonalization
of the molecular Hamiltonian over the basis set of the electronic-rotational and nuclear spin wave functions. Details of the
method can be found in
Refs. \cite{Petrov2013, Skripnikov2015}.

\section{\label{sec:Anal}Data and Analysis}
\begin{figure}[ht]
 \centering
 \includegraphics[width=8cm]{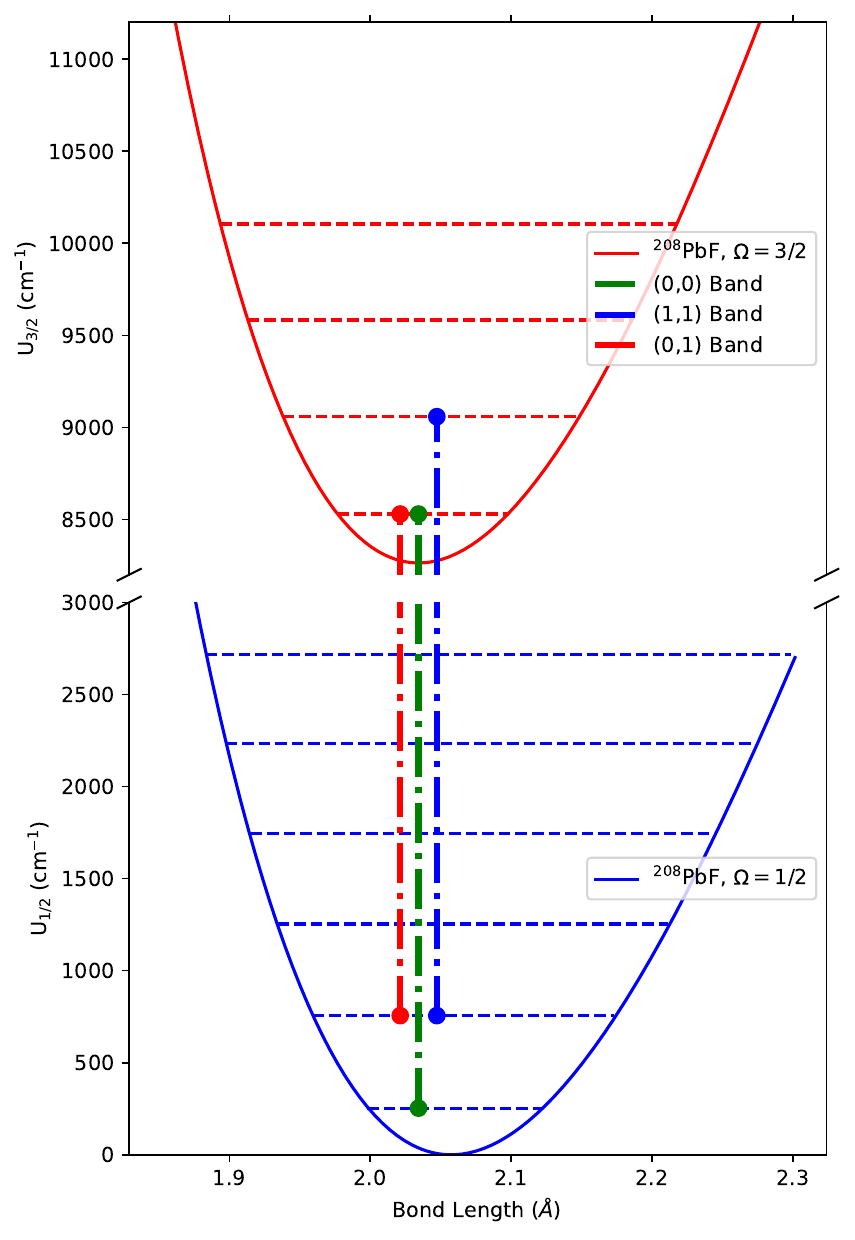}
  \caption{\label{fig:NIRTransitions}\small{Scheme of the fine structure transitions measured for $^{206}$PbF, $^{207}$PbF and $^{208}$PbF. The potential curves and level positions are calculated from the present results }}
\end{figure}

In order to characterize fully the mixed $X_1$/ $X_2$ $^2\Pi$ electronic state complex of PbF, we have combined the 130 measured microwave transition frequencies with the data published by Ziebarth et al. \cite{Ziebarth1991, Ziebarth1998} who measured spectroscopic emission transitions between the $X_2\,^2\Pi_{3/2}$ and $X_1\,^2\Pi_{1/2}$ fine structure states in the near-IR (NIR) region at 1.3 $\mu m$.  Figure \ref{fig:NIRTransitions} illustrates the vibronic transitions involved.  Each of those shown in the figure includes many rotationally-resolved, and for \PbF\, hyperfine-resolved, transitions and there are data for the 3 most abundant naturally-occurring isotopic variants of the molecule.  In addition to the transitions originally identified, \cite{Ziebarth1991, Ziebarth1998} we have assigned many transitions in the \textit{v} = (1,1) band in $^{208}$PbF and \PbF. We are most grateful to Prof. E. Fink (Wuppertal) who generously provided us with a copy of the original data and additional details.  The complete set of 1201 NIR and 130 FTMW spectroscopic transitions included in the present analysis is available in Jackson's thesis \cite{Jacksonthesis}.\\

The analysis used the {\sc {\sc spfit}} code of the {\sc calpgm} package originally developed by Pickett \cite{Pickett1991, Novick2016} for the purpose of categorizing astronomical line positions which has become a ubiquitous \textit{de facto} standard in the field.  Employing control code sequences, the elements comprising the Hamiltonian matrix are highly adaptable and can handle spectroscopic data of many types and precision, including multiple isotopic variants of a molecule in multiple vibrational levels with fine and nuclear hyperfine interactions. This allows the use of a Dunham-type expansion \cite{Shirts2018} to describe the molecular energy levels assuming standard reduced mass variation of the parameters.  Details are given in the original description of the {\sc calpgm} suite of programs by Pickett \cite{Pickett1991} and a guide by Novick \cite{Novick2016}. Dunham coefficients, $Y_{l,m}$, derived from fitting to experimental data can be related to parameters describing the diatomic potential function \cite{GordyandCook}. More details are given in the sections  below.  \\

A published analysis of data for the two bromine isotopes in BrO \cite{Drouin2001} illustrates many of the capabilities of the {\sc {\sc spfit}} package that are relevant to the present work.  The form of the rotational energy terms, with the equilibrium rotational constant, $B_e$, Dunham coefficient $Y_{01}$, $\Lambda$-doubling parameter, $p_{00}$, and hyperfine parameters that were expressed in the Frosch and Foley representation \cite{Frosch1952} are discussed in detail by Brown and Carrington \cite{Brown2003}.  A brief overview is given below in section \ref{subsec:DunhamForm}.  Finally, small contributions from the nuclear spin-rotation coupling, $C_I$, for both $^{19}$F and $^{207}$Pb and the nuclear spin-spin dipolar coupling interaction, $t_{00}$, for \PbF\,, were found to be needed for the reproduction of the spectra at the accuracy and precision inherent to the FTMW data.

\section{\label{sec:RandD}Results and Discussion}
\subsection{\label{subsec:ResultsIntro}Overall Fitting}
The FTMW data give very precise information on rotational, $\Lambda$-doubling and hyperfine spacings in the $X_1$ component, but has little or no sensitivity to the larger fine-structure or vibrational spacings that define the position and properties of the upper $X_2$ component. On the other hand, the NIR data determine parameters describing the fine structure and vibrational spacings and also contain information on centrifugal distortion effects because the data extend to high rotational quantum number.  Lower precision measurements for the upper $X_2$ component hyperfine and $\Lambda$-doubling splittings meant that some parameters involving them were not determinable.  The spin-orbit splitting parameter $A_{e}$ determines the spacing between the $X_2$ and $X_1$ states and is normally expected to be weakly vibrationally dependent and isotopically independent, however the data indicated these assumptions were invalid and the variation of the spin-orbit splitting was accounted for in a similar way to that adopted by Drouin et al. \cite{Drouin2001} in the case of BrO.  More details are given in the following sub-sections. The numerical results are presented in Tables \ref{tab:MajorParms}, \ref{tab:PbHFS} and \ref{tab:FHFS}. 

\begin{table} 
\begin{center}
\caption{{\small Vibrational, fine-structure and rotational parameters derived from a global PbF fit of 1331 transitions (weighted fit $\sigma$ = 1.02; see text). Parameters without parenthetical error values were kept at values calculated prior to the fit, see text for details. Isotopic scaling relations are from \cite{Drouin2001}.}} 
\label{tab:MajorParms}
\begin{ruledtabular}
\resizebox{0.9\textwidth}{!}{
\begin{tabular}{c r r}
 & \multicolumn{1}{l}{\begin{tabular}[c]{@{}c@{}}$^{208}$PbF Molecular Parameters\end{tabular}} & \multicolumn{1}{r}{\begin{tabular}[c]{@{}c@{}}Isotopic Scaling Relations\end{tabular}} \\

Parameter & Value\footnotemark[1] & Reduced Mass Ratio
\\

\textit{Y$_{10}$} & 15593865.4(13) & $\mu^{-1/2}$ \\

\textit{Y$_{20}$} \footnotemark[3] & -66070(14) & $\mu^{-1}$ \\
 
\textit{Y$_{01}$} & 6937.14652(11) & $\mu^{-1}$ \\

\textit{Y$_{01}$} \textit{(BOB)} & -2.520(26) kHz & $\delta \langle r^2 \rangle$ \\

$V_{Pb}$ \footnotemark[2] & 48.8(4) ($10^{-7}$ fm$^{-2}$) & none \\

\textit{Y$_{11}$} & -44.10637(11) & $\mu^{-3/2}$ \\

\textit{Y$_{21}$} & 0.080705(38) & $\mu^{-2}$ \\

\textit{Y$_{31}$} & 0.1713(37) kHz & $\mu^{-5/2}$ \\
 
\textit{Y$_{02}$} & -5.4844(72) kHz & $\mu^{-2}$ \\

\textit{A$_{e}$} & 247705376.9(46) & none \\

$\gamma_{00}$ & -2513.6(44) & $\mu^{-1}$ \\

\textit{Y$^{*}$$_{00}$} & 5647.2 & $\mu^{-1}$\\

$^{207}$PbF \textit{$\Delta A_{e}^{iso}$} & 199.2(21) & none \\

$^{206}$PbF \textit{$\Delta A_{e}^{iso}$} & 312.0(22) & none\\

\textit{A$_{10}$} & 828312.4(56) & $\mu^{-1/2}$ \\

\textit{A$_{20}$} \footnotemark[3] & -23708(15) & $\mu^{-1}$ \\

\textit{A$_{01}$} & 156.9284(57) & $\mu^{-1}$ \\

\textit{A$_{11}$} & 1.0223(50) & $\mu^{-3/2}$ \\

\textit{A$_{02}$} & 0.1804(13) kHz & $\mu^{-2}$ \\

\textit{p$_{00}$} & -4142.46361(25) & $\mu^{-1}$ \\

\textit{p$_{00}$ (BOB)} & -5.240(63) kHz & $M_{ref}-M_{\alpha}$ \\

$\Delta ^{p_{00}}_{01}$ \footnotemark[2] & 99.4(96) (unitless) & none \\

\textit{p$_{10}$} & -2.29053(25) & $\mu^{-3/2}$ \\

\textit{p$_{20}$} & 0.238350(43) & $\mu^{-2}$ \\

\textit{p$_{01}$} & -3.125(19) kHz & $\mu^{-2}$ \\

\end{tabular}}
\end{ruledtabular}
\SetSinglespace{1.1}\singlespacing
\begin{singlespace*} 
    \footnotetext[1]{In MHz unless otherwise noted, 1$\sigma$ deviation given in parenthesis.}
    \footnotetext[2]{The field shift, $V_{Pb}$, and the Born-Oppenheimer Breakdown (BOB) term, $\Delta^{p_{00}}_{01}$, were not determined from the fit directly. Fitted parameters that correspond to the offset values used in {\sc spfit} program are $Y_{01}$ BOB and $p_{00}$ BOB. The quoted uncertainties for $V_{Pb}$ and $\Delta^{p_{00}}_{01}$ were derived using the uncertainties of the fitted parameters and Equations \ref{Eqn14} and \ref{Eqn16}.}
    \footnotetext[3]{Uncertainty in this parameter comes from a preliminary fit of the NIR data alone. The $Y_{10}$ and $Y_{20}$ parameters give the average of the harmonic and anharmonic vibrational constants between the $X_1$ and $X_2$ states. These parameters relate to $Y^*_{lm}$ as in Equation \ref{eqn:Ystar}.  The $\omega_ex_e$ = Y20$^*$ values reported in \cite{Ziebarth1991} equal the Y20 (average) value here.}

\end{singlespace*}

\end{center}
%\label{tab:MajorParms}
\end{table}

 \begin{table}[ht]
 \begin{center}
      \caption{$^{207}$Pb hyperfine parameters derived from the global fit, as in Table \ref{tab:MajorParms}}  
      \label{tab:PbHFS}
     \begin{ruledtabular}
     \resizebox{0.9\textwidth}{!}{
     \begin{tabular}{c c c}
     
Parameter & Value\footnotemark[1] & Reduced Mass Ratio
\\

\textit{a$_{00(Pb)}$} \footnotemark[2] & 2775(40) & g$_N$ \\

\textit{b$_{F_{00(Pb)}}$} & -2207(140) & g$_N$ \\

\textit{b$_{F_{10(Pb)}}$} & -27.4121(49) & g$_N \mu^{-1/2}$ \\

\textit{b$_{F_{20(Pb)}}$} & 0.1257(17) & g$_N \mu^{-1}$ \\

\textit{c$_{00(Pb)}$} \footnotemark[2] & -3584(200) & g$_N$ \\

\textit{d$_{00(Pb)}$} & 7246.9964(22) & g$_N$ \\

\textit{d$_{10(Pb)}$} & 34.0789(49) & g$_N \mu^{-1/2}$ \\

\textit{d$_{20(Pb)}$} & -28.6(19) kHz & g$_N \mu^{-1}$ \\

\textit{d$_{01(Pb)}$} & 7.01(12) kHz & g$_N \mu^{-1}$  \\

\textit{C$_{I_{00(Pb)}}$} & 0.07793(28) & g$_N \mu^{-1}$ \\

\textit{C$_{I_{10(Pb)}}$} & 0.77(33) kHz & g$_N \mu^{-3/2}$\\

\textit{t$_{00}$} & -3.41(78) kHz & g$_N$ \\

Parameter \footnotemark[3] & Theory Value  & Experimental Value  \\

\textit{A$_{\parallel}\, (\frac{1}{2})$} & 9849, 9796\footnotemark[5] & 10146.6733(9)\footnotemark[4]\\

\textit{A$_{\perp}$} & -6990, -6911\footnotemark[5] & -7264.0287(25), -7264.0388(4)\footnotemark[4] \\

\textit{A$_{\parallel}\, (\frac{3}{2})$} & ... & 318(50) 
\\

\textit{\~{A}$_{\perp}$(=b$_{00(Pb)})$} & -1012(120)\footnotemark[2], -1217\footnotemark[5] & ... \\

     \end{tabular}}
     \end{ruledtabular}

     \SetSinglespace{1.1}\singlespacing
\begin{singlespace*} 
      \footnotetext[1]{In MHz unless otherwise noted, 1$\sigma$ deviation given in parenthesis.}
      \footnotetext[2]{Determined iteratively from experiment ($a_{00\,(Pb)}$) and a combination of experiment and theory ($c_{00\,(Pb)}$). $c_{00\,(Pb)}$ uses the $b_{00\,(Pb)}$ theoretical value. See Section \ref{subsec:hyperfine} for details, both were fixed in the final fit.}
      \footnotetext[3]{Alternative formulation of hyperfine terms. These parameters are not determined from the fit directly, rather they are calculated from the optimized Frosch-Foley $a$, $b_F$, $c$, and $d$ parameters. The relationships are given in the text in Section \ref{subsec:hyperfine}.}
      \footnotetext[4] {Reference \cite{Mawhorter2011}}
      \footnotetext[5] {Reference \cite{Petrov2013}}
 \end{singlespace*}   
 \end{center}
 \end{table}

 \begin{table}[ht]
 \begin{center}
      \caption{$^{19}$F hyperfine parameters derived from the global fit, as in Table \ref{tab:MajorParms}}  
      \label{tab:FHFS}
     \begin{ruledtabular}
     \begin{tabular}{c c c}
Parameter & Value\footnotemark[1] & Reduced Mass Ratio \\

\textit{a$_{00(F)}$} & 129(5) & g$_N$ \\

\textit{b$_{F_{00(F)}}$} & 49.2418(12) & g$_N$ \\

\textit{b$_{F_{10(F)}}$} & 3.0314(13) & g$_N \mu^{-1/2}$ \\

\textit{b$_{F_{20(F)}}$} & -21.09(26) kHz & g$_N \mu^{-1}$ \\

\textit{c$_{00(F)}$} & -308(17) & g$_N$ \\

\textit{d$_{00(F)}$} & -255.03763(58) & g$_N$ \\

\textit{d$_{10(F)}$} & -1.91608(71) & g$_N \mu^{-1/2}$ \\

\textit{d$_{20(F)}$} & 20.35(16) kHz & g$_N \mu^{-1}$ \\

\textit{d$_{01(F)}$} & -0.536(66) kHz & g$_N \mu^{-1}$\\

\textit{C$_{I_{00(F)}}$} & 4.401(79) kHz & g$_N \mu^{-1}$ \\

Parameter \footnotemark[2] & Theory Value \footnotemark[2] & Experimental Value\footnotemark[2] \\

\textit{A$_{\parallel}\,(\frac{1}{2})$} & 412(11) & 409.8416(14)  \\

\textit{A$_{\perp}$} & 254(2) & 255.9906(7)  \\

\textit{A$_{\parallel}\,(\frac{3}{2})$} & 34.3(21) & ... \\

\textit{\~{A}$_{\perp}$(=b$_{00(F)})$} & 153.0(92) & ... \\

     \end{tabular}
     \end{ruledtabular}
\SetSinglespace{1.1}\singlespacing
\begin{singlespace*} 
        \footnotetext[1]{In MHz unless otherwise noted, 1$\sigma$ deviation given in parenthesis.}
        \footnotetext[2]{These parameters are not determined from the fit directly, rather they are calculated from the optimized Frosch-Foley $a$, $b_F$, $c$, and $d$ parameters. The relationships are given in the text in Section \ref{subsec:hyperfine}.}
\end{singlespace*}
\end{center}
 \end{table}

\subsection{\label{subsec:DunhamForm}Dunham Formulation}
The global multi-isotope fit of the data to a Dunham-type expansion comprises 45 independent parameters, most of which describe the energy levels of a diatomic molecule within a series of products of powers of a vibrational quantum number-dependent term, $(v+1/2)^l$, and a rotational term as $(N(N+1))^m$.  Each operator term is preceded by a coefficient $X_{l,m}$, that is to be determined by a least squares minimization fitting the data. Parameter choices were guided by those used by Cohen et al. in their study \cite{Cohen2006} of isoelectronic BiO which included vibrational states up to \textit{v} = 9. \\ 

To ensure the final parameters for the $X_1$/ $X_2$ $^2\Pi$ electronic state complex also reproduce vibronic level spacings to higher vibrational quantum number derived from lower resolution spectra reported by Lumley and Barrow \cite{Lumley1977,Huber1979} and the initial Ziebarth et al. \cite{Ziebarth1991} data, a preliminary fit of all the NIR data including the vibronic spacings from Ref. \cite{Ziebarth1991} to determine the harmonic and anharmonic vibrational constants $Y_{10}\,\textrm{and}\,Y_{20}$ and their corresponding $A_{lm}$ corrections was performed.  The global fit including the FTMW data was then performed with $Y_{20}$ fixed and the two steps repeated to ensure convergence.  Below, we summarize some of the essential attributes of the Dunham expansion model as implemented here. \\

The rotational and vibrational energy levels can be calculated as described in \cite{GordyandCook}: 
\begin{equation} \label{eqn:RotDunham}
    \frac{E_{vN}}{h} = \sum_{l,m} Y_{l,m}(v+\frac{1}{2})^{l}N^{m}(N+1)^{m}
\end{equation}
\myremm{Here, the $Y_{lm}$ are the Dunham coeffeicients and we have adopted} the case (b) \cite{Brown2003} definition of the rotational angular momentum, $\mathbf{N}\,=\, \mathbf{J} - \mathbf{S}$, as used in the {\sc spfit} code.  The first few terms in the rotation-vibration energy, for example, are given by 
\begin{equation} \label{eqn:RotVib}
\begin{split}
   \frac{E_{vN}}{h}=Y_{10}(v+\frac{1}{2})+Y_{20}(v+\frac{1}{2})^{2}+Y_{01}N(N+1)+Y_{11}(v+\frac{1}{2})N(N+1)+\\Y_{21}(v+\frac{1}{2})^{2}N(N+1)+Y_{31}(v+\frac{1}{2})^{3}N(N+1)+Y_{02}N^{2}(N+1)^{2}+Y_{03}N^{3}(N+1)^{3}
\end{split}
\end{equation}
Additional terms may be added as needed to adequately fit the data. Other angular momentum operator-based terms representing $\Lambda$-doubling and nuclear spin hyperfine interactions were added in a similar fashion \cite{Drouin2001}.   \\

\myremm{A complication in PbF comes from the fact that the  $X_1$ and $X_2$ fine structure components in the molecule have different potentials and therefore exhibit different effective vibrational frequencies.}  This difference is handled by defining a \myremm{modified} set of Dunham coefficients \myremm{($Y_{lm}^*$)} for each fine-structure component:
\begin{equation}\label{eqn:Ystar}
    Y^*_{lm}\,=\, Y_{lm} \pm A_{lm}/2.
\end{equation}
where $Y_{lm}$ are the standard Dunham coefficients, \myremm{ Eqns. (\ref{eqn:RotDunham},\ref{eqn:RotVib}), that may be thought of as representing the average potential} and the upper and lower signs in (\ref{eqn:Ystar}) refer to the $X_2$ and $X_1$ component respectively. \myremm{The $A_{lm}$ are the fine structure energy contributions derived from the coefficient of the base operator term $A_e L_z S_z$ in the Hamiltonian, defined analogously to the $Y_{lm}$:}  
\begin{equation} \label{eqn:SpinOrbit}
    A_{v}=A_{00}+A_{10}(v+\frac{1}{2})+A_{20}(v+\frac{1}{2})^{2}+...
\end{equation}
and, \myremm{to account for additional non Born-Oppenheimer subtleties discussed below and in section \ref{subsec:BOBEffects}:}
\begin{equation}\label{eqn:Aiso}
    A_{00}^{iso}\,=\,A_e + \left(Y_{00}^*(^2\Pi_{3/2}) - Y_{00}^*(^2\Pi_{1/2})\right) + \Delta A_e^{iso},
\end{equation}
for the lower $X_1$ and upper $X_2$ component vibrational frequencies and anharmonicities in cm$^{-1}$ units.   Note that the various spin-orbit $A$ parameters appearing in the equations above include $\Delta A_e^{iso}$, an additional isotopic variation not normally encountered, due here to a Born-Oppenheimer breakdown effect due to the variation in Pb-nuclear size.  More discussion of this effect is included in section \ref{subsec:BOBEffects}.  We find $\omega_e$($X_1$, $X_2$) = (506.341(1), 533.970(1)) cm$^{-1}$ and $\omega_ex_e$($X_1$, $X_2$) = (2.599(1), 1.809(1)) cm$^{-1}$, in reasonable agreement with earlier band head measurements in \cite{Huber1979}, \cite{BrownWatson1977} and \cite{Ziebarth1991}, as well as recent calculations \cite{Luan_2024}, taking into account the differences in experimental resolution and basis set dependence, respectively. \\

To obtain the experimental fine structure spacing we need to add $-\gamma$ to agree with the numbers in reference \cite{Mawhorter2011}, see subsection \ref{subsec:ADGamma}.  The difference between the upper and lower component potentials is obvious in Figure \ref{fig:NIRTransitions} where the potential curves were computed using potential parameters derived from the final Dunham expansion parameters in Table \ref{tab:MajorParms} and the relationship between the Dunham coefficients and potential coefficients\,\cite{GordyandCook}.\\

In the limit where the Born-Oppenheimer approximation holds, the constant $A_e$ is isotope-independent\cite{Drouin2001}, but this was not found here for PbF. The parameters in (\ref{eqn:Aiso}), $Y^{*}_{00}\,=\,\left(Y_{00}^*(^2\Pi_{3/2}) - Y_{00}^*(^2\Pi_{1/2})\right)$ and $\Delta A^{iso}_e$ account for the observed deviations. The $Y^{*}_{00}$ parameter is a correction term describing the difference between $A_{00}$ and $A_e$, or the difference between the (experimentally measured) fine structure interval for the $v=0$ levels and the equilibrium fine structure spacing of the $X_2$ and $X_1$ $^{2}\Pi$ components.  In the multi-isotope fit, $A_e$ is for the reference $^{208}$Pb isotope. The $Y^{*}_{00}$  parameters for \PbF\, and $^{206}$PbF  are isotopically scaled from the $^{208}$PbF $Y^{*}_{00}$ value by $\mu^{-1}$\cite{Drouin2001}. While these terms should account for the isotopic dependence of $A_{00}$ leading to a constant $A_e$, as is the case for BrO \cite{Drouin2001} there remained (small) systematic deviations between the $A_e$ spacing in the different isotopologues and these deviations are the $\Delta$$A_e^{iso}$ parameters in the results.  No isotopic scaling relation is assumed for the $\Delta$$A_e^{iso}$ values; they are simply constant correction terms of the order of .0001\% of the value of $A_e$ corresponding to the $\approx$100 MHz nonlinearity in the isotopic dependence of $A$ already observed by Ziebarth et al. \cite{Ziebarth1998}. In subsection \ref{subsec:BOBEffects}, they are interpreted as energy contributions due to finite-sized nucleus effects. \\

The Dunham formulated parameters relate to conventional rotational and centrifugal distortions parameters as 
\begin{equation}
    B_{v}=Y_{01}+Y_{11}(v+\frac{1}{2})+Y_{21}(v+\frac{1}{2})^{2}+Y_{31}(v+\frac{1}{2})^{3}+...
\end{equation}
for the effective rotational constant, and
\begin{equation}
    D_e\,=\,-Y_{02}
\end{equation}
and
\begin{equation}
    H_e\,=\,Y_{03}
\end{equation}
for the centrifugal distortion constants. \\

\subsection{\label{subsec:hyperfine}Hyperfine Parameters}

For a $^2\Pi$ molecule like PbF, Frosch and Foley \cite{Frosch1952} showed there are, in principle, four hyperfine coupling terms that need to be specified for each nuclear magnetic moment in the molecule.  The terms in the effective Hamiltonian have coefficients conventionally labeled $a$, $b$, $c$ and $d$.  The first represents the coupling between the electronic orbital angular momentum projection, $\Lambda$, and the nuclear spin angular momentum, while the remaining three terms can be cast as combinations of the zero and second rank spherical tensor operators representing the coupling between the unpaired electron spin and the nuclear spin angular momenta.  Brown and Carrington \cite{Brown2003} detail the operator forms and show how they are related to classical dipole-dipole coupling terms.\\

From the electron spin resonance (ESR) literature the tensor nomenclature $A_{||}(1/2)$, $A_{||}(3/2)$, $A_{\perp}$ and $\tilde{A}_{\perp}$ has also been utilized \cite{Petrov2013, Mawhorter2011}. The parallel quantities represent linear combinations of Frosch and Foley terms entering in the diagonal matrix elements for the $X_1\,^2\Pi_{1/2}$ and $X_2\,^2\Pi_{3/2}$ wave-functions and are related to the Frosch and Foley coefficients by $A_{||}(1/2) = (2a-b-c)$ and $A_{||}(3/2) = (2a+b+c)/3$. Note that Brown and Carrington introduce the analogous quantities $h_{1/2} = A_{||}(1/2)/2$ and $h_{3/2} = 3A_{||}(3/2)/2$.  The physically important Fermi Contact interaction is $b_F = b+c/3$, while the angular dependence of the electron-nuclear spin dipolar interaction is contained in the constants $c$ and $d$. The latter is equal to $-A_{\perp}$ and determines the parity variation in the coupling of the spin dipole-dipole coupling while  the constant $c$ is related to the the $Y_{20}$ spherical harmonic distribution of the electronic wavefunction \cite{Brown2003}. Finally, $\tilde{A}_{\perp}\, =\, b$. \\

The FTMW data determine the combinations $(2a - (b+c))$, i.e. $A_{||}(1/2)$, and $d$, i.e.$-A_{\perp}$ for the $^{207}$Pb nucleus in $^{207}$PbF. When the $^{207}$PbF $X_2 - X_1$ NIR data were included in the analysis some additional flexibility in the model was found to be needed in order to model the data to the measurement precision and accuracy.  After some trials, we combined information from theory to pin the value of $\tilde{A}_{\perp} (=b)$ and optimized the Frosch and Foley $a$ parameter to the observed $^{207}$PbF NIR data while forcing the combination of parameters $A_{||}(1/2)$ at the value of 10146.7 MHz obtained previously \cite{Mawhorter2011,Petrov2013} from the FTMW \textit{v} = 0 data.  The complete dataset was then fit to determine a corresponding value for $b_F$ as well as $d$.  Note that $d$ = -$A_{\perp}$ is functionally independent of the other hyperfine parameters since it only determines the difference in hyperfine splittings between levels of opposite parity. These steps were repeated until convergence. In this way, we obtain estimates for all the Frosch and Foley parameters which are given, with their statistical uncertainties, in Table \ref{tab:PbHFS}. A similar procedure was used to determine the analogous $^{19}$F hyperfine parameters given in Table \ref{tab:FHFS}.

\subsection{\label{subsec:ADGamma}Separating $A_{01}$ and $\gamma_{00}$}

$A_{01}$, the centrifugal correction to the spin-orbit coupling and $\gamma_{00}$, the electron spin-rotation coupling, appearing as  $\gamma_{00}(\mathbf{N}\cdot\mathbf{S})$ in the Hamiltonian, have indistinguishable effects on the energies \cite{BrownWatson1977}.  The physical interactions described by these parameters are distinct, but their contributions to the energy levels of the molecule are not, which makes them difficult to determine separately using spectroscopic data. Brown and Watson \cite{BrownWatson1977} showed that $A_{01}$, or $A_D$ in their notation, and $\gamma_{00}$ may be separated by using data from multiple isotopic variants of the molecule and individually determined from the equation: 
\begin{equation}\label{eqn:AD_gamma}
    \frac{\tilde{A}_{01}}{Y_{01}} \,=\, \frac{A_{01}}{Y_{01}} - \left[ \frac{2Y_{01}}{A_e - 2Y_{01}} \right] \left( \frac{\gamma_{00}}{Y_{01}}\right).
\end{equation}
Here, $\tilde{A}_{01}$ is the effective $A_{01}$ parameter determined when $\gamma_{00}$ is held at zero in a fit to spectroscopic data for a single isotopic variant of a molecule. The actual $A_{01}/Y_{01}$ and $\gamma_{00}/Y_{01}$ ratios are isotope independent; they have identical isotopic mass ratios.  The true values of $A_{01}$ and $\gamma_{00}$ can therefore be determined by using data for multiple isotopes to form simultaneous equations (\ref{eqn:AD_gamma}) and solving them for the constants. \\

$\gamma_{00}$ enters linearly in the energy expression for the $X_1$ component \cite{Brown2003} and it is therefore also very strongly correlated with $A_e$.  Within the experimental measurement precision, the FTMW data are not sensitive to $A_e$ at all, so only the NIR transitions were used in a procedure to separate the spin-rotation and centrifugal distortion to the spin-orbit coupling. An initial value for $\gamma_{00}$ was estimated from second-order perturbation theory. Here $\gamma_{00}$ = $(A_{00}^{iso} \times Y_{01})/\Delta E$ assuming a model where only a single $\,^2\Sigma^+$ state of PbF \cite{McRaven2010} at an excitation energy of 22 500 \wn\,  contaminates the lowest $X_1$ component of the ground $^2\Pi$ state \cite{Brown2003}. Then, we estimate that the second-order contribution to $\gamma_{00}$ is 
\begin{equation}\label{eqn:gamma2}
    \gamma^{(2)}_{00}\,=\, -\frac{A_e \times Y_{01}}{2.25\times 10^{4}} \,\approx \, -0.078\, \mathrm{cm}^{-1}\,\, \mathrm{or}\, -2300\,\mathrm{MHz}.
\end{equation}

The NIR data for all isotopes of PbF were then fit together assuming $A_{01}$ and $\gamma_{00}$ obey their known isotopic mass ratio dependencies, as in Table \ref{tab:MajorParms}, thus forcing the ratios of $A_{01}/Y_{01}$ and $\gamma_{00}/Y_{01}$ to remain constant for each isotopologue. In this way, we enforce the relationship between the two parameters in equation (\ref{eqn:AD_gamma}), and determine both a $\gamma$ parameter and an $A_{01}$ parameter simultaneously.  The value for $A_e$ was adjusted iteratively to match the fine structure splitting until convergence.  The final value determined for $\gamma_{00}$ of -2513.6(44) MHz (see Table \ref{tab:MajorParms}) is close to that estimated from equation (\ref{eqn:gamma2}). \\

Due to the indistinguishable nature of $A_{01}$ and $\gamma_{00}$ previous analyses have not utilized a $\gamma$ parameter.  Including the nonzero $\gamma_{00}$ affects the $X_2 - X_1$ spin-orbit splitting as it enters linearly in the $X_1$ energy.  With $\gamma_{00}$ included, we find values of {248 116 740}, {248 117 020}, and {248 117 216} MHz for the splitting in $^{208}$PbF, $^{207}$PbF and $^{206}$PbF respectively, in excellent agreement with the multiple values in \cite{Mawhorter2011}. \\

The final parameters reproduce the observed spectroscopic data to within the measurement uncertainty. The overall standard deviation of the multi-isotope fit was 1.02 relative to the measurement uncertainties, reflecting a very satisfactory result.  The resulting parameters may be used to predict the positions of unobserved levels and, in particular, to estimate the splitting between the low-lying, opposite parity levels of interest in \PbF \,.  These spacings were not measured directly but the derived parameters from our analysis permits their estimation.  The results are given in column 5 of Table \ref{tab:VibEngLife}, where they are also compared to the computed spacings based on the \textit{ab initio} results.  The experimental results show that the spacing is a minimum of 15.1 MHz with an uncertainty of less than 100 kHz, in the excited vibrational level, \textit{v} = 8. \\

  \begin{table}[ht]
  \begin{center}
      \caption{Vibrational energies, lifetimes in the $X_1\,^2\Pi_{1/2}$ state of $^{208}$PbF and the $J$=$\left( \frac{1}{2}\right)^{+}$, $F\,=\,\frac{1}{2}$, and $J$=$\left( \frac{1}{2}\right)^{-}$, $F\,=\,\frac{1}{2}$ parity level splitting in $^{207}$PbF. } \label{tab:VibEngLife}
     \begin{ruledtabular}
     \begin{tabular}{c c c c c c}
       Vibrational & Experimental\footnotemark[1]  & Calculated \footnotemark[2] & Lifetime\footnotemark[2] & \multicolumn{2}{c}{$\pm$\, Parity State Spacing / MHz} \\
             Level, \textit{v} &   cm$^{-1}$  &   cm$^{-1}$ & msec & Experiment\footnotemark[1] & Calculated \footnotemark[2]   \\
  0 &    0.0     &     0.0    &  \footnotemark[3]  & 266.285(2) &   266.3  \\
  1 &  502.72    &   502.2    &   277(28)          & 233.505(5) &   232.8 \\
  2 &  1000.89   &   999.8    &   141(15)          & 200.037(9) &   199.3 \\
  3 &  1494.49   &  1492.9    &    95(10)          & 165.881(15)&   164.9 \\
  4 &  1983.54   &  1981.5    &    72(7)           & 131.039(24)&   129.9 \\
  5 &  2468.02   &  2465.6    &    59(6)           & 95.513(34) &   94.0  \\
  6 &  2947.94   &  2945.3    &    49(5)           & 59.304(47) &   57.4  \\
  7 &  3423.31   &  3420.7    &    43(4)           & 22.415(61) &   20.4 \\
  8 &  3894.11   &  3891.6    &    38(4)           &-15.152(78) &   -17.1  \\
  9 &  4360.36   &  4358.3    &    34(3)           &-53.394(97) &  -47.4 \\
       
     \end{tabular}
     \end{ruledtabular}
      \footnotetext[1]{Calculated using the fit parameters and their associated errors} from Tables \ref{tab:MajorParms}, \ref{tab:PbHFS} and \ref{tab:FHFS}.  Where shown, one standard deviation uncertainty is given in parenthesis in units of the last quoted decimal place.
     \footnotetext[2]{CCSD(T) method.  Computational details are given in Section \ref{sec:Theory}}.
     \footnotetext[3]{This is the lowest vibronic level.}

     \end{center}
 \end{table}

\subsection{\label{subsec:BOBEffects}Field Shifts and Born-Oppenheimer Breakdown Effects}

 Molecular field shift effects were first observed in the seminal 1982 studies of related lead chalcogenide and thallium halide diatomics \cite{Schlembach1982, Tiemann1982}.\,  Based on ideas developed to explain field shifts in atomic spectra, Schlembach and Tiemann \cite{Schlembach1982} showed how changes in the Coulomb potential experienced by the electrons in a molecule due to the finite size of nuclei can be accounted for.  The imposition of isotopic mass-scaling relationships between the $Y_{lm}$ in the present analysis resulted in small (few kHz) systematic errors in the fit residuals due to the different isotopic field shifts. Deviations of this type from conventional spectroscopic models are known as Born-Oppenheimer breakdown (BOB) effects and are one manifestation of a larger set of consequences of the Born-Oppenheimer separation of electronic and nuclear motion \cite{Watson1980} that is implicitly assumed in the use of a Dunham expansion to describe the molecular energies.\\

 More recent theoretical studies of field shift effects on rotational \cite{Knecht2012} and vibrational \cite{Almoukhalalati2016} spectra for PbTe, TlI, PtSi and the original PbX and TlX molecules, respectively, give summaries of the topic and also present new relativistic calculations. Almoukhalalati et al. \cite{Almoukhalalati2016} also include the 2007 study of PbO \cite{Serafin2007} in their analysis.  Knecht and Saue \cite{Knecht2012} pointed out a factor of $\pi^2$ ($\approx 10$) inconsistency in a 1985 follow-up study by the Tiemann group \cite{Knockel1985} which Prof. Tiemann has recently confirmed \cite{TiemannPC}. Hence their field shift quantities quoted below are from the original 1982 study by Schlembach and Tiemann. Other examples have been noted in our recent work on AlCl, BiF, and BiCl \cite{Preston2022} as well as BaF \cite{Preston2024}. \\

In fact, we need to allow for these effects in not only the rotational constants, $Y_{01}$, but also in the parameters for $\Lambda-$doubling, $p_{00}$, and the spin-orbit coupling, $A_e$ (as noted in subsection \ref{subsec:DunhamForm}) as described by Watson \cite{Watson1980} and Brown and Carrington \cite{Brown2003}. Fluorine has only one stable isotope, so for PbF the $Y_{01}$ BOB term on isotopic substitution of nucleus $X'$ for reference nucleus $X$ is given in the Schlembach-modified Watson model by \cite{Serafin2007}
\begin{equation}\label{eq:Y01massdep}
    Y_{01}\,=\,\mu^{-1}\overline{U}_{01}\left[ 1 + m_e\frac{\Delta ^X_{01}}{M_{X'}} + V_X \delta\langle r^2 \rangle_{X X'}\right]
\end{equation}
where $\mu$ is the reduced mass, $m_e$ is the electron mass, $M_{X'}$ is the atomic mass of the substituted isotope, $\Delta_{01}$ is a Watson-type mass-dependent BOB correction and $V_X\delta\langle r^2 \rangle_{XX'}$ is the field shift term that depends on the change in mean square charge radii between the two nuclei $X$ and $X'$.  The $\overline{U}_{01}$ term is:

\begin{equation}
    \overline{U}_{01}\,=\, U_{01}\left( 1 + V_X\langle r^2 \rangle_X\right)
\end{equation}
where $U_{01}$ is the isotope-independent rotational spectral parameter, $U_{01} = \mu Y_{01}$, and the $\langle r^2 \rangle$ term is the mean square charge radius of the reference nucleus $X$. Further discussion of mean square radii changes and a table of values for $\delta\langle r^2 \rangle$ are given in \cite{Angeli2013}.\\

In the fitting, these correction terms were implemented as additive corrections to the isotopically scaled $Y_{01}$ parameter for each isotopologue, $Y_{01}^{\alpha}$.  $^{208}$PbF is the reference isotopologue, and field shift corrections were added to the isotopically scaled values of $Y_{01}$ for the other three isotopologues. We can then calculate the value of the field-shift correction term $V_{Pb}$ as
\begin{equation}
    Y_{01}^{\alpha}+(FSC)=\mu_{\alpha}^{-1}\overline{U}_{01}[1+V_{Pb}\delta\langle r^2 \rangle_{Pb,Pb'}],
\end{equation}
where the $FSC$ term is the required field shift correction determined from the fitting, and the $\alpha$ index denotes different isotopologues of PbF. This equation can be rearranged for $V_{Pb}$ as 
\begin{equation}
\label{Eqn14}
    V_{Pb}=\Bigl{(}\frac{Y_{01}^{\alpha}+(FSC)}{\mu^{-1}_{\alpha}\overline{U}_{01}}-1\Bigl{)}/\delta <{r^2}>_{Pb,Pb'}.
\end{equation}

Using the values determined for the field shift correction, $Y_{01}^{\alpha}$, and $\overline{U}_{01}$, we find $V_{Pb}=48.8(4)$ x $10^{-7}$ fm$^{-2}$.\\

For most high-Z atoms like Pb and Tl discussed here, the heaviest
isotope is the most abundant. This generally determines the reference
isotopologue, as is the case here with $^{208}$PbF. Hence successive
isotopologues in equation (11) move from heaviest to lightest and both
the $\Delta_{01}^X$ and $V_X$ terms are consistent with
experimentally observed negative $Y_{01}$ offsets. This is the case because
(1): $\Delta_{01}^X$ is negative and the lighter isotope masses in the denominator of the mass-dependent term make this term more negative in an
essentially linear fashion, and (2): although $V_X$ is positive, progression to lighter nuclei generally leads to an increasingly negative trend in $\delta\langle r^2 \rangle$ and thus in the overall nuclear size dependent field shift term as well.\\

\begin{figure}[ht]
    \centering
    \includegraphics[scale=4.85]{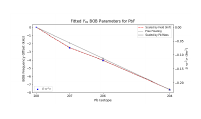}
    \caption{Plotted frequency offsets to $Y_{01}$ vs. Pb isotope for three different cases. The black dotted line shows the BOB frequency offsets for each isotopologue when determined separately in the fit and free to move to the best fit value. Note the excellent agreement of the free-floating BOB frequency offsets and those scaled by $\delta\langle r^2 \rangle$ values, opposed to those scaled by the Pb isotope mass. The mean square radii changes from the $^{208}$PbF reference isotopologue shown in blue are from \cite{Angeli2013}. Errors are the same size as point markers. }
    \label{fig:BOBGraph}
\end{figure}

The difference between the $\Delta_{01}^X$ and $V_X$ terms is that the individual
$\delta\langle r^2 \rangle$ values lend physically significant shape to
the $V_X$ negative trend. This helps explain the generally better fits in
the literature to the residuals using only the field shift term (i.e. $\Delta_{01}^X\equiv 0$) than when using only the more rigid linear mass-dependent term
(i.e. $V_X \equiv 0$). Of course this is particularly true when the field shift
term is much larger than the mass dependent term, which we expect to be the case for PbF. Indeed, Figure \ref{fig:BOBGraph} shows that the residuals follow the shape and trend of the $\delta\langle r^2 \rangle$ values almost exactly. Numerically, a one parameter $V_X$ fit has an RMS of 1.0197, very close to the minimum RMS of 1.0195 with all 4 residuals floating independently. In contrast, for a rigidly linear one parameter mass-dependent fit this quantity grows significantly worse to 1.0421. \\

As noted by previous authors, the resulting very strong
correlation between these terms in many cases precludes cleanly
untangling them. In some cases, e.g BaF \cite{Preston2024}, a
strong odd-even variation in $\delta\langle r^2 \rangle$ can help
significantly, but that is not the case for Pb or Tl (which has
only 2 stable isotopes). The $\Delta_{01}^X$ term itself is made up of three smaller Dunham, adiabatic, and nonadiabatic terms \cite{Schlembach1982,Serafin2007}. As is the case in the detailed discussion for PbO, it usually requires some assumptions to determine all 3. For an open shell molecule like PbF, the presence of the unpaired electron prevents reliably estimating the rotational $g-$factor, $g_J$, in the nonadiabatic term, which makes up 95\% of the overall $\Delta_{01}^{Pb}$ term for PbO \cite{Serafin2007}.  Another approach is to calculate the field shift term. The illuminating study by Knecht and Saue\cite{Knecht2012} shows that
non-relativistic calculations are completely inadequate, and that
extensive relativistic calculations provide agreement with experiment
values at the 10-20\% level. \\

Nevertheless, we can use the multiple alternate fits reported with
either $\Delta_{01}^{Pb} \equiv 0$ \cite{Schlembach1982} or $V_X \equiv 0$ \cite{Giuliano2008, Tiemann1982, Serafin2007} to draw a fairly clear
picture of the relative importance of these terms. Table V summarizes $V_X$
values from these studies for PbY (Y = Se, Te, S, O) and TlY (Y = F, Cl,
Br, I) alongside our current PbF result. The more recent $V_{Pb}$  results\cite{Giuliano2008} for PbSe and PbTe agree nicely, and estimates ranging from 10\% to 16\% for the relative fraction of $\Delta_{01}^{Pb}$ character in the 4 lead chalcogenides are also included. These have been derived by noting the consistent and very large factor of 6 to 10 jumps in the $\Delta_{01}^{Pb}$ term when the field shift term is excluded in studies of PbSe and PbTe \cite{Giuliano2008},
PbS \cite{Tiemann1982, Schlembach1982} and PbO \cite{Serafin2007}. Units of 1/fm$^2$ are used to match the fm$^2$ units of $\delta\langle r^2 \rangle$.  The alternative approach is to exclude the mass-dependent term, which Schlembach and Tiemann used in their original study for PbSe, PbTe, and PbS as well as for TlF, TlCl, TlBr, and TlI \cite{Schlembach1982}. They found that this limiting case results only in changes in $V_X$ of less than 2$\sigma$ for each PbY molecule and less than 1$\sigma$ for the whole TlY set. \\

The amount of mass-dependent character is smaller for PbS and PbO than
for PbSe and PbTe, and this can be understood both in terms of the
smaller sizes of S and O atoms as well as their increasing electronegativity due to the physical dependence of $V_X$ on the derivative $d\rho/dr$ of the charge density at the nucleus \cite{Serafin2007, Giuliano2008}. For PbF, fluorine is small and even more electronegative, and indeed our reported $V_{Pb}$ value of 48.8(20) x 10$^{-7}$ fm$^{-2}$ is closer to that found for the more ionic TlY molecules, where it ranges from 32.0 to 40.9 x 10$^{-7}$ fm$^{-2}$ from TlI to TlF. Our PbF value assumes $\Delta_{01}^{Pb} \equiv 0$ and the alternate $V_{Pb}$ only fit shown in Figure 3 validates this assumption. While we are unable to reliably estimate it, we are confident $\Delta_{01}^{Pb}$  is quite small. To estimate an uncertainty, based on Schlembach and Tiemann's experience with $\Delta_{01}^X \equiv 0$ analyses \cite{Schlembach1982}, we use their $1\sigma$ value of 1.9 x 10$^{-7}$ fm$^{-2}$ for the most similar case of TlF and add it in quadrature with our much smaller statistical uncertainty of 0.4 x 10$^{-7}$ fm$^{-2}$.

\begin{table}[ht]
\begin{center}
\caption{Born-Oppenheimer breakdown field shift parameters for Pb and Tl diatomic molecules.}
        \label{tab:BOB}
    \begin{ruledtabular}
    \begin{tabular}{c c c c}  
    
     \textbf{Molecule} & \textbf{V}$_X$ ($10^-7$ fm$^{-2}$)\footnotemark[1]   & \textbf{  $\Delta_{01}^{Pb}$ character} & Reference\\

      $^{208}\mathrm{Pb}^{80}\mathrm{Se}$ & 22.1(19) & - & \cite{Schlembach1982}\\
       $ $ & 22.5(20) & 0.15 & \cite{Giuliano2008}\\

        $^{208}\mathrm{Pb}^{130}\mathrm{Te}$ & 21.2(16) & - & \cite{Schlembach1982}\\
        
        $ $ & 20.7(21) & 0.16 & \cite{Giuliano2008}\\
 
        $^{208}\mathrm{Pb}^{32}$S & 24.5(19) & - & \cite{Schlembach1982}\\

        $ $ & " & 0.10 & \cite{Schlembach1982, Tiemann1982}\\

         $^{208}\mathrm{Pb}^{16}$O & 26.4(36) & 0.13 & \cite{Serafin2007}\\

          $^{208}\mathrm{Pb}^{19}$F & 48.8(20)\footnotemark[2] & - & this work\\
 
        $^{205}\mathrm{Tl}^{19}$F & 40.9(19) & -  & \cite{Schlembach1982}\\
         
        $^{205}\mathrm{Tl}^{35}$Cl & 40.9(55) & - & \cite{Schlembach1982}\\

        $^{205}\mathrm{Tl}^{79}$Br & 33.7(10) & - & \cite{Schlembach1982}\\

        $^{205}\mathrm{Tl}^{127}$I & 32.0(10) & - & \cite{Schlembach1982}\\
    
    \end{tabular}
    \end{ruledtabular}
    
    \footnotetext[1]{All uncertainties are 1$\sigma$ unless otherwise noted.}
    \footnotetext[2]{Assuming $\Delta_{01}^{Pb} = 0$. See text for details of the estimated uncertainty.}
    \end{center}
\end{table}

The $\Delta A_e$ parameter in Table \ref{tab:MajorParms} may be interpreted as the contribution to the electronic transition energy between $X_1$ and $X_2$ resulting from the extended size of the nucleus. If this is the case, then $d \Delta A_e / d\langle r^2 \rangle$ should be similar to the field shift factor $F=d \nu / d\langle r^2 \rangle$ (where $\nu$ is the atomic transition energy) commonly used in atomic isotope shift studies. For atoms, this parameter is considered to be constant, i.e. independent of the isotope pair.
By referring to Table \ref{tab:MajorParms} and Ref. \cite{Knecht2012}, we may calculate that 
%$d \Delta A_e / d\langle r^2 \rangle (208-207{\rm Pb})$ = -2810 MHz/fm$^2$, 
$d \Delta A_e / d\langle r^2 \rangle (208-207{\rm Pb})$ = -2681(28) MHz/fm$^2$, 
and 
%$d \Delta A_e / d\langle r^2 \rangle (208-206{\rm Pb})$ = -2668 MHz/fm$^2$. 
$d \Delta A_e / d\langle r^2 \rangle (208-206{\rm Pb})$ = -2624(19) MHz/fm$^2$;
this seems very reasonable agreement.  \\

For the $\Lambda$-doubling parameter $p_{00}$, second-order perturbation theory also shows that $p_{00}$ is proportional to $(A_{00}^{iso} \times Y_{01})/\Delta E$, with an extra factor of 2 compared with $\gamma_{00}$ \cite{Brown2003}. Indeed $p_{00}$ $\approx$ -4140 MHz is roughly a factor of 2 greater than the earlier $\gamma_{00}$ estimate of -2300 MHz. Furthermore, our values of $p_{00}/(A_{00}^{iso} \times Y_{01})$ are quite constant, pointing to the overall consistency of this analysis. \\

Based on the above arguments, the nonlinear isotope dependence of $A_{00}^{iso}$ will also be present in the $p_{00}$ parameter and can be accounted for using a mass-dependent $\Delta_{01}$ BOB parameter, determined following the method described by Preston \cite{Prestonthesis}. Using the relations 
\begin{equation}
\label{Eqn15}
    \delta^{p_{00}}_{01}=\Big(\frac{p_{00}^{\alpha}+\text{offset}}{\mu^{ref}/\mu^{\alpha}}-p_{00}^{ref}\Big)\Big/\frac{M^{\alpha}-M^{ref}}{M^{\alpha}}
\end{equation} and 
\begin{equation}
\label{Eqn16}
    \Delta^{p_{00}}_{01}=-\delta_{01}\Big(\frac{M^{ref}}{m_e}\Big)\Big[(p_{00}^{ref}+\delta_{01})^{-1}\Big],
\end{equation} 

where the ``offset'' term is the frequency offset to the $p_{00}$ term determined in the fitting, and $^{208}$Pb$^{19}$F is used as the reference isotopologue. We find a unitless Watson-type $\Delta^{p_{00}}_{01}$ value of 99(10). This is large compared to typical values for rotational constants, but we are not aware of other published values of $\Delta^{p_{00}}_{01}$ for comparison. We also note that the BOB frequency offsets for $V_{Pb}$ and $\Delta^{p_{00}}_{01}$ are found to be of comparable size at about $-2.5$ kHz and $-5$ kHz respectively, while as a percentage the $p_{00}$ BOB frequency offset is about another factor of 2 larger.

\subsection{Sensitivity of the PbF molecule to variation of fundamental constants}

This section and the next deal with the special sensitivity of PbF to symmetry-breaking effects. As already noted, Luan et al. \cite{Luan_2024} have further investigated the feasibility as well as some practical considerations of laser cooling the lead halides PbX (X = F, Cl, Br, I).  They used relativistic methods and find quite diagonal Frank-Condon factors (FCFs) in each case.  These greatly facilitate the laser cooling required to study subtle symmetry-breaking effects, and PbF is the lightest of the halides possessing the smallest level density. \\

Nowadays the possibility of observing the space-time variation of the fine-structure constant $\alpha=e^2/\hbar c$, the electron-to-proton mass ratio, $m_e/m_p$, and the dimensionless fundamental parameter for strong interactions $m_q/\Lambda_\mathrm{QCD}$, where $m_q$ is the light quark mass and $\Lambda_\mathrm{QCD}$ is the QCD scale, is of great interest. The parameter $m_q/\Lambda_\mathrm{QCD}$ enters atomic physics through nuclear magnetic $g$-factors \cite{Flambaum04,Tedesco06}.  To date, almost all searches for space-time variation in fundamental constants have been performed on atoms, the only exception being the experiment on the rovibrational transitions of the SF$_6$ molecule \cite{Shelk08}. \\

In  Ref. \cite{Flambaum:2013} it was noted that  the existence of closely spaced levels of opposite parity also enhances the sensitivity of molecular spectra to variation of fundamental constants $\alpha$ and $m_q/\Lambda_\mathrm{QCD}$.
For variation of the frequency $\omega = E(F^{p}=1/2^{+}) - E(F^{p}=1/2^{-}) = 266.285$ MHz for $v=0$
with respect to $\alpha$ and $(m_q/\Lambda_{QCD})$ it was found
\begin{eqnarray}
\frac {\delta \omega} {\omega} \approx -55 \frac{\delta \alpha} {\alpha} + 2.1 \frac {\delta (m_q/\Lambda_{QCD}) }{(m_q/\Lambda_{QCD})}.
\label{var}
\end{eqnarray}

One can see from Eq. (\ref{var}) that the $^{207}$Pb$^{19}$F molecular radical species can offer about two orders of magnitude (two orders means that the coefficient in front of the  $\frac{\delta \alpha} {\alpha}$ in Eq. (\ref{var}) is of order $10^2$).) enhancement of the relative effect of
$\alpha$-variation. This is comparable to the enhancements in some other molecular species \cite{Flambaum06,FK07b,DSS08,ZKY08}. However, the sensitivity coefficient for $m_{q}/\Lambda_{\mathrm{QCD}}$ is enhanced by two orders of magnitude compared with the ratio of frequencies of $^{133}$Cs and $^{87}$Rb atomic clocks, which currently provide the best limit on the variation of $m_{q}/\Lambda_{\mathrm{QCD}}$ \cite{Guena12}. \\

In general the variation of the frequency $\omega = E(F^{p}=1/2^{+}) - E(F^{p}=1/2^{-}) $ 
with respect to $\alpha$ and $(m_q/\Lambda_{QCD})$ can be written as
\begin{equation}
\label{der1}
\frac{\partial \omega} {\partial \alpha} = \frac{\partial \omega} {\partial p} \frac{\partial p} {\partial  \alpha}
+ \frac{\partial \omega} {\partial A^{Pb}_{\parallel}} \frac{\partial A^{Pb}_{\parallel}} {\partial  \alpha}
+\frac{\partial \omega} {\partial A^{Pb}_{\perp}} \frac{\partial A^{Pb}_{\perp}} {\partial  \alpha}, 
\end{equation}
\begin{align}
\frac{\partial \omega} {\partial (m_q/\Lambda_{QCD}) } &= 
\label{der2}
 \frac{\partial \omega} {\partial A^{Pb}_{\parallel}} \frac{\partial A^{Pb}_{\parallel}} {\partial (m_q/\Lambda_{QCD})}
+\frac{\partial \omega} {\partial A^{Pb}_{\perp}} \frac{\partial A^{Pb}_{\perp}} {\partial  (m_q/\Lambda_{QCD})}. 
\end{align}

For the $X_1$ component, the $A_{\parallel}$ and $A_{\perp}$ parameters are defined in section \ref{subsec:hyperfine}.  From the results of Flambaum et al. \cite{Flambaum:2013}, we can show

\begin{align}
\label{dirF}
 \frac{\partial p} {\partial  \alpha} &= 2.1538 p / \alpha, \\
 \frac{\partial A^{Pb}_{\parallel}} {\partial  \alpha} &= 4.39 A^{Pb}_{\parallel} / \alpha, \\
 \frac{\partial A^{Pb}_{\perp}} {\partial  \alpha} &= 4.39  A^{Pb}_{\perp} / \alpha, \\
 \frac{\partial A^{Pb}_{\parallel}} {\partial (m_q/\Lambda_{QCD}) } &= -0.111 A^{Pb}_{\parallel} / (m_q/\Lambda_{QCD}), \\
 \frac{\partial A^{Pb}_{\perp}} {\partial (m_q/\Lambda_{QCD}) } &= -0.111 A^{Pb}_{\perp} / (m_q/\Lambda_{QCD}).
\end{align}

According to the present calculations: 
\begin{align}
\frac{\partial \omega} {\partial p} &= -0.838, \\
\frac{\partial \omega} {\partial A^{Pb}_{\parallel}} &= -0.153, \\
\frac{\partial \omega} {\partial A^{Pb}_{\perp}} &= +0.507,
\label{dirv7}
\end{align}
for $v=7$ and:
\begin{align}
\frac{\partial \omega} {\partial p} &= +0.835, \\
\frac{\partial \omega} {\partial A^{Pb}_{\parallel}} &= +0.154, \\
\frac{\partial \omega} {\partial A^{Pb}_{\perp}} &= -0.509.
\label{dirv8}
\end{align}
for $v=8$.

Combining Eqs. (\ref{dirF})-(\ref{dirv8}) we have
\begin{equation}
\frac {\delta \omega} {\omega} \approx -720 \frac{\delta \alpha} {\alpha} + 27 \frac {\delta (m_q/\Lambda_{QCD}) }{(m_q/\Lambda_{QCD})}
\label{var7}
\end{equation}
for $v=7$ and
\begin{equation}
\frac {\delta \omega} {\omega} \approx 1084 \frac{\delta \alpha} {\alpha} - 39 \frac {\delta (m_q/\Lambda_{QCD}) }{(m_q/\Lambda_{QCD})}
\label{var8}
\end{equation}
for $v=8$.

One can see from Eqs. (\ref{var}, \ref{var7}, \ref{var8}) that the $v=7$ and $v=8$ vibrational levels offer a further enhancement  about one order of magnitude higher as compared to the $v=0$ level. 

We note, that the corresponding absolute values for the frequencies shifts
\begin{equation}
\frac {\delta \omega} {\rm GHz} \approx -15 \frac{\delta \alpha} {\alpha} + 0.56 \frac {\delta (m_q/\Lambda_{QCD}) }{(m_q/\Lambda_{QCD})}
\label{vara}
\end{equation}
for $v=0$,
\begin{equation}
\frac {\delta \omega} {\rm GHz} \approx -16 \frac{\delta \alpha} {\alpha} + 0.61 \frac {\delta (m_q/\Lambda_{QCD}) }{(m_q/\Lambda_{QCD})}
\label{var7a}
\end{equation}
for $v=7$ and
\begin{equation}
\frac {\delta \omega} {\rm GHz} \approx -16 \frac{\delta \alpha} {\alpha} + 0.59 \frac {\delta (m_q/\Lambda_{QCD}) }{(m_q/\Lambda_{QCD})}
\label{var8a}
\end{equation}
for $v=8$ are about the same (Note (see Table \ref{tab:VibEngLife}), that the frequency $\omega$ for $v=8$ is assumed to be negative ). 
From Eqs. (\ref{var7},\ref{var8}) one can see that the sensitivity coefficients for the $v=7$ and $v=8$ levels have opposite signs that could be a powerful tool for suppressing possible systematic errors.
%Moreover, the sensitivity coefficients for the $v=7$ and $v=8$ levels have opposite signs that could be used for suppressing possible systematic errors.
Coefficients in fronts of $\frac{\delta \alpha} {\alpha}$ and
$\frac {\delta (m_q/\Lambda_{QCD}) }{(m_q/\Lambda_{QCD})}$ scale very similarly for different transitions.
Therefore, unfortunately, it is difficult for an experiment to differentiate between variations of fundamental constants using only the PbF molecule.

\subsection{Sensitivity of the PbF molecule to the electron electric dipole moment}

In a polar molecule the $\mathcal{T,P}$-violating energy shift associated with $e$EDM reads
\begin{equation}
\Delta E_{\mathcal{P},\mathcal{T}}= P E_{\rm eff}  d_e,
\label{shift}
\end{equation}
where $d_e$ is the value of the $e$EDM, $E_{\rm eff}$ is effective electric field determined by the electronic structure of the molecule, and $P$ is the corresponding ${\mathcal{P},\mathcal{T}}$-odd polarization coefficient. In Ref. \cite{Skripnikov:14c} we calculated that $E_{\rm eff} = 40$ GV/cm.\\

It is well known that for $\Omega=1$ diatomics (such as HfF$^+$ and ThO), due to the existence of the $\Omega$-doublet structure, the $P$ value approaches unity for small ($\sim 10-100$ V/cm) laboratory electric fields. In general, for $\Omega=1/2$ diatomics a much larger electric field is required because the parity doublets are further apart in energy. For example for the YbF molecule in the ground $^2\Sigma_{1/2}$ state $P$ is only $0.55$ for $E=10$ kV/cm \cite{Tarbutt2013}. For $^{206, 208}$PbF (spin-less even isotopes of Pb) at $E=5$ kV/cm the polarisation is about 80\%  \cite{Baturo2021}. \\

Due to the accidental near cancellation of the $\Omega$-type doubling and magnetic hyperfine interaction energy shifts, the situation is different for $^{207}$PbF. In this case the energy spacing between levels of opposite parity is much smaller than the typical value for $\Omega=1/2$ diatomics. The latter makes the molecule more promising for the $e$EDM  search experiment.  Figure 1 of \cite{Mawhorter2011} shows the  hyperfine split levels of interest for $v=0$. The energy levels of interest for an $e$EDM  search experiment on $^{207}$PbF are those from the second to the fifth close $\Omega$-doublets levels $F^{p}=3/2^{-}$, $F^{p}=1/2^{-}$, $F^{p}=1/2^{+}$ and $F^{p}=3/2^{+}$ associated with $J=1/2$ in the figure. \\
 
As calculated in Ref. \cite{Baturo2021} \PbF\, as compared to $^{206, 208}$PbF can be polarized by a smaller electric field, which is an advantage since larger electric field tends to lead to larger systematic effects. For example for $^{207}$PbF $F=3/2$, $|M_F|$=3/2, $v=0$, $P=0.8$ polarization is achieved for  $E=1$ kV/cm. For $E=2$ kV/cm, $P$ is about 90\%.\\

\begin{figure*}[t!]
%    \centering
%    \begin{subfigure}[t]{0.4\textwidth}
%        \centering
        \includegraphics[height=6.0cm]{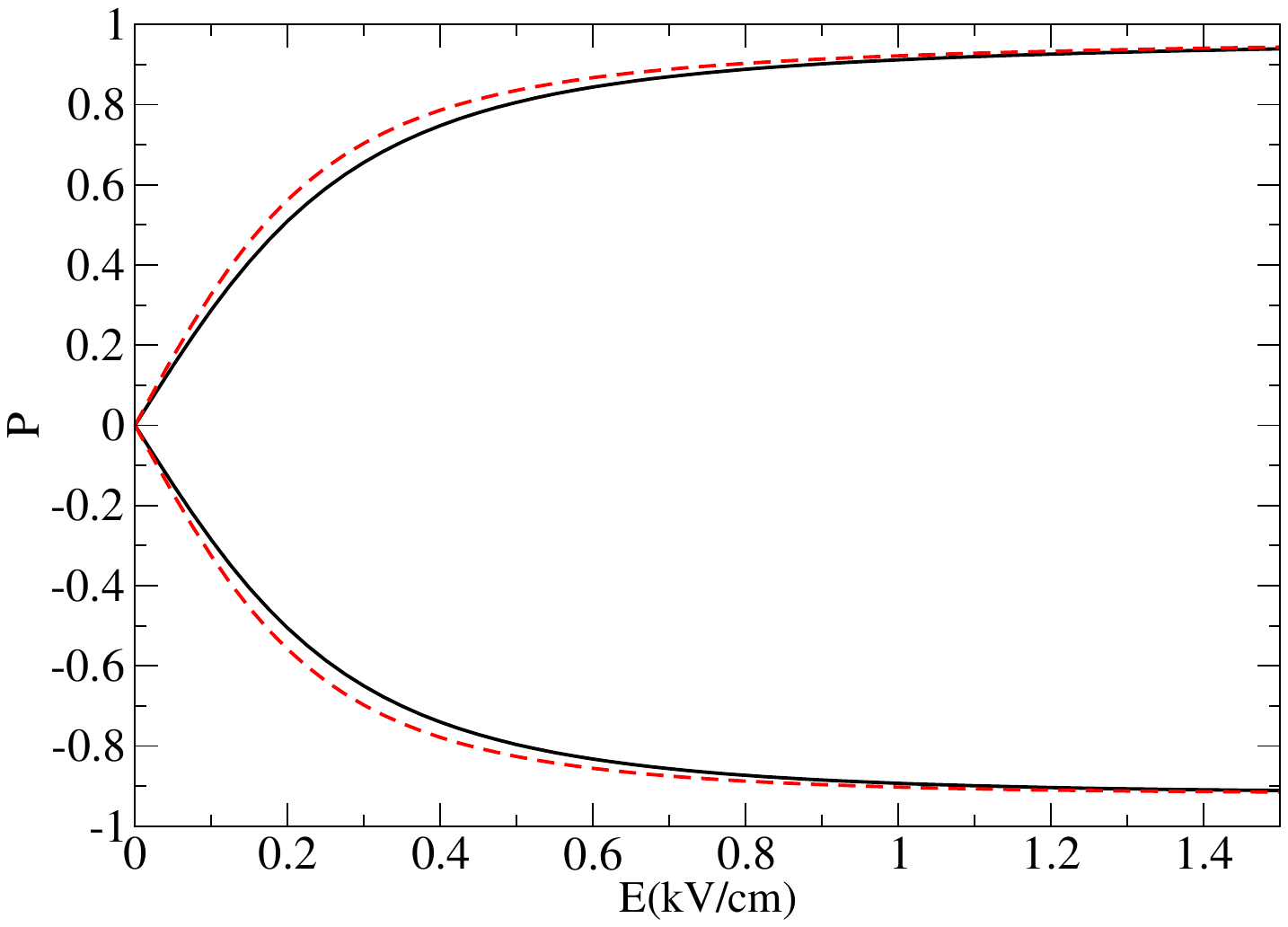}
        \includegraphics[height=6.0cm]{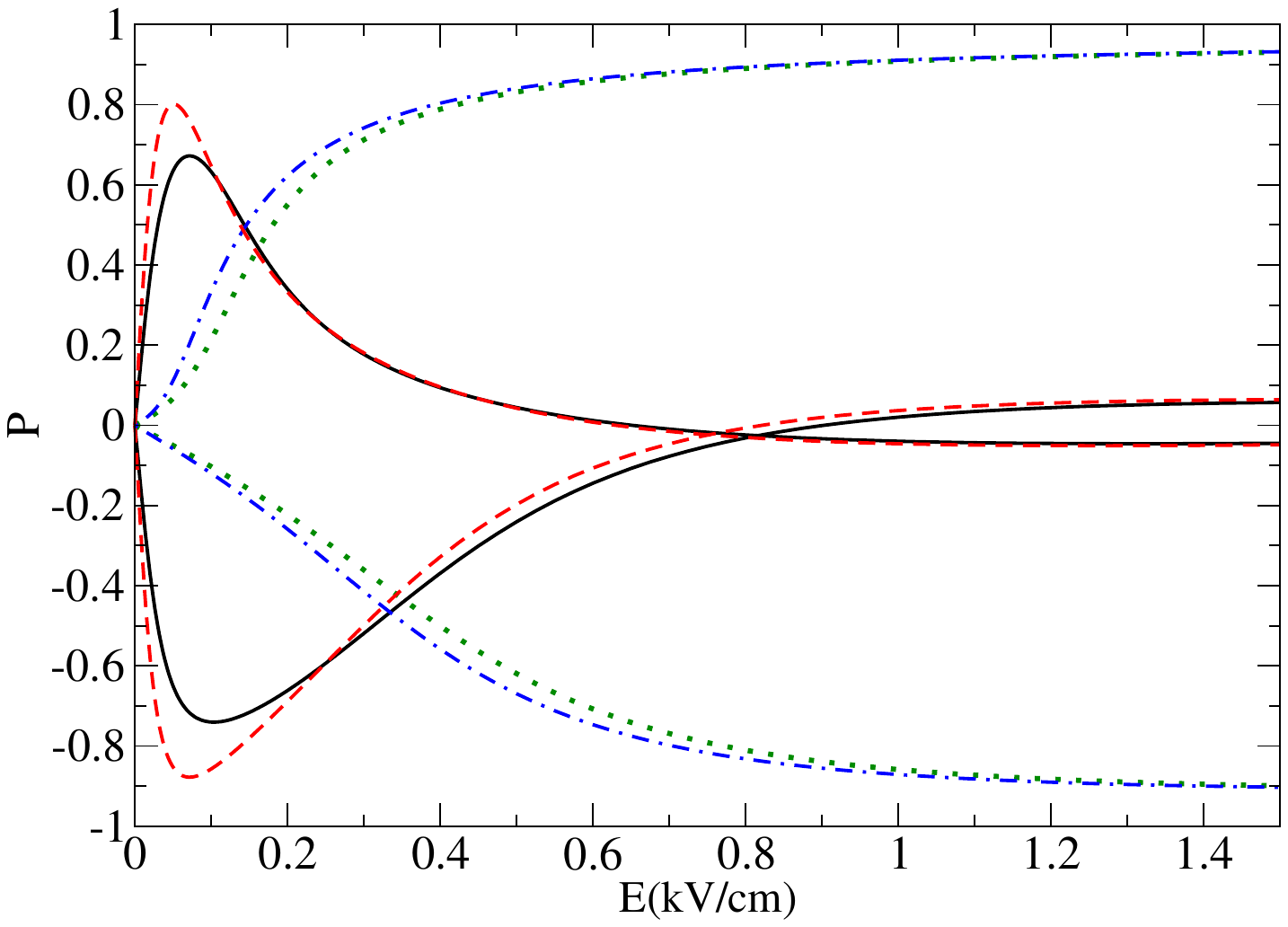}
%        \caption{\small{The transition in the \textit{v} = 3 level.}}
%    \end{subfigure}
    \caption{\label{fig:P32andP12}
Left: Calculated polarization $P$ as function of the electric field $E$ for $F=3/2$ $M_F=3/2$.
 Solid (black) lines correspond to  $v=7$, 
 dashed (red) lines correspond to $v=8$ vibrational levels.
 $\Omega$-doublets levels marked by the same colour. Right: Calculated polarization $P$ as function of the electric field $E$.
 Solid (black) lines correspond to $F=1/2$ $M_F=1/2$ $v=7$, 
 dashed (red) lines correspond to $F=1/2$ $M_F=1/2$ $v=8$,
 dotted (green) lines correspond to $F=3/2$ $M_F=1/2$ $v=7$,
 dashed-dotted (blue) lines correspond to $F=3/2$ $M_F=1/2$ $v=8$.
 $\Omega$-doublets levels marked by the same colour.
 }
\end{figure*}

In Fig. \ref{fig:P32andP12} the calculated polarization $P$ for $v=7$ and $v=8$ are shown.  For $^{207}$PbF $F=3/2$, $|M_F|$=3/2, $P=0.9$ is achieved for $E=1$ kV/cm for both $v=7$ and $v=8$ levels. For $^{207}$PbF $F=1/2$, $|M_F|$=1/2, $v=8$ we already have $P=-0.9$ for an electric field less than 100 V/cm.

\section{Conclusions}

Lead monofluoride is a strong molecular candidate for current and future investigations of the role of parity non-conservation in theories beyond the standard model, as well as in hunting for the extra parity non-conservation, PNC, effects needed to explain the observed fundamental matter-antimatter baryon asymmetry in the universe.  We also show that PbF may be a useful vehicle for investigating possible time variation of the fundamental constants and a better understanding of the spin-dependent nuclear PNC effect known as the anapole moment. \\

The present results offer a great deal of information needed for further investigation of these effects, for example complementing the several ongoing studies of the eEDM.  The ground state sensitivity of \PbF\, to PNC effects is optimized by the near-degeneracy of opposite parity states in the $v\,=\,8$ level, where they are found to be only 15 MHz apart. Importantly, we find that the lifetime of this level is 38 ms, long enough for proposed optical cooling experiments \cite{Luan_2024} to be employed and also longer than the typical transit time limitations in beam measurements of buffer gas-cooled samples \cite{Hutzler2012}. The isotopically independent parameters describing the coupled $X_1\, / \, X_2$ states of the molecule determined here have resulted in a much more accurate picture of their properties.  Highlights of the results include:

\begin{itemize}
 \item The new high resolution $X_2,v=1$ - $X_1, v=1$ NIR data and an improved model to account for isotopic shifts has resulted in a much clearer picture of the $X_1$ and $X_2$ spin-orbit split molecular potentials.  
 \item Born-Oppenheimer breakdown (BOB) effects in both the rotation constant $Y_{01}$ and the lambda-doubling constant $p_{00}$ have been observed and accounted for. A simple single perturbing state model finds that $p$ is proportional to $(A \times Y_{01})$, and this closely mimics the results.  Hence the isotopic dependence observed for $p_{00}$ can be traced back to that in $A_e$ and the $X_1$ - $X_2$ energy spacing discussed above.
 \item The rich isotopic data set has enabled a separation of the effects of the  centrifugal distortion correction to the spin-orbit term $A_{01}$ and the electron spin-nuclear rotation interaction $\gamma$ that usually result in indistinguishable contributions to the energy.
 \item A comparison with other PbX and TlY BOB studies showed that essentially all of the $Y_{01}$ BOB effect is due to the finite nuclear size of Pb, and the measured residuals match the trend in nuclear sizes almost exactly.
 \item The case for the sensitivity of PbF to the variation of fundamental constants has been augmented as well as the primary eEDM sensitivity noted above.
\end{itemize}
Future work could entail further excited state characterization based on laser spectroscopy of the $X_2-X_1$ band system in a cooled sample which could resolve remaining ambiguities in the hyperfine splittings in the upper component, as well as work to experimentally test the laser slowing theory posited recently \cite{Luan_2024}.  Both would be a good further steps towards realizing the full potential of PbF for molecular PNC investigations and possible studies of the time variation of fundamental constants put forward here.

\section{Acknowledgements}
The Pomona College authors thank fellow student Jos\'e Mu$\tilde{n}$oz-Lopez for assistance taking data. They also appreciate research support provided by a Pomona College Sontag Fellowship as well as Hirsch Research Initiation and Summer Undergraduate Research Program grants. Ewald Fink and Klaus Setzer from Wuppertal are acknowledged for kindly sharing their unpublished $v = 1-1$ data.  TJS acknowledges support by the U.S. Department of Energy, Office of Science, Division of Chemical Sciences, Geosciences and Biosciences within the Office of Basic Energy Sciences, under Award Number DE-SC0018950 for part of this work. JUG acknowledges support from the Deutsche Forschungsgemeinschaft (DFG) under GR1344/11-1 and from the Land Niedersachsen. Calculations of sensitivity to variation of fundamental constants and to electron electric dipole moment are supported by the Russian Science Foundation grant no. 24-12-00092.

\newpage
%\bibliography{References, PetrovLib}% Produces the bibliography via BibTeX.
%\begin{thebibliography}{99}
\providecommand{\noopsort}[1]{}\providecommand{\singleletter}[1]{#1}%
%

%\end{thebibliography}

\newpage
\appendix 
% \addcontentsline{toc}{chapter}{Appendices}
\section{}

% \chapter{FTMW Transitions of PbF}
The following table lists the observed FTMW lines of PbF used in the global fit. The quantum numbers listed appear as they do in the SPFIT program, meaning the $J$ quantum numbers are rounded up to the nearest integer number in this table. Additionally, the isotopologues of PbF are distinguished via the $v$ quantum number. For $^{207}$PbF, $v=0$ is listed as $v=10$, $v=1$ is $v=11$, and so on. For $^{206}$PbF, $v=0$ is listed as $v=20$, $v=1$ is $v=21$, and so on. For $^{204}$PbF, $v=0$ is listed as $v=30$. 

\begin{longtable}[h]{||c c c c c c c c c c c c c c ||}

 \hline
 \multicolumn{14}{| c |}{\textbf{FTMW Transitions of PbF}}\\
 \hline
$N'$ &	$P'$&	$v'$&	$J'$&	$F_1'$&	$F'$&	$N''$&	$P''$&	$v''$&	$J''$&	$F_1''$&	$F''$&	Measured Freq (MHz). &	Unc. \\
 \hline
 \endfirsthead

 \hline
 \multicolumn{14}{|c|}{\textbf{FTMW Transitions of PbF}}\\
 \hline
$N'$ &	$P'$&	$v'$&	$J'$&	$F_1'$&	$F'$&	$N''$&	$P''$&	$v''$&	$J''$&	$F_1''$&	$F''$&	Measured Freq (MHz). &	Unc. \\
 \hline
 \endhead

 \hline
 \endfoot

 \hline
 \multicolumn{14}{|c|}{End of Table}\\
 \hline\hline
 \endlastfoot

1&	1&	0&	1&	0&	0&	1&	-1&	0&	1&	1&	1&	3922.50648&	0.002\\
1&	1&	0&	1&	1&	1&	1&	-1&	0&	1&	0&	0&	4194.77734&	0.0007\\
1&	1&	0&	1&	1&	1&	1&	-1&	0&	1&	1&	1&	4229.71764&	0.003\\
2&	-1&	0&	2&	1&	1&	2&	1&	0&	2&	1&	1&	8117.30169&	0.001\\
2&	-1&	0&	2&	1&	1&	2&	1&	0&	2&	2&	2&	8199.8478&	0.0009\\
2&	-1&	0&	2&	2&	2&	2&	1&	0&	2&	1&	1&	8307.51802&	0.002\\
2&	-1&	0&	2&	2&	2&	2&	1&	0&	2&	2&	2&	8390.06637&	0.002\\
3&	1&	0&	3&	2&	2&	3&	-1&	0&	3&	2&	2&	12277.68224&	0.0007\\
3&	1&	0&	3&	2&	2&	3&	-1&	0&	3&	3&	3&	12374.67007&	0.0007\\
3&	1&	0&	3&	3&	3&	3&	-1&	0&	3&	2&	2&	12443.85768&	0.0007\\
3&	1&	0&	3&	3&	3&	3&	-1&	0&	3&	3&	3&	12540.84651&	0.0008\\
4&	-1&	0&	4&	3&	3&	4&	1&	0&	4&	3&	3&	16428.51597&	0.001\\
4&	-1&	0&	4&	4&	4&	4&	1&	0&	4&	4&	4&	16688.49294&	0.002\\
2&	1&	0&	2&	2&	4&	1&	-1&	0&	1&	1&	1&	18414.58796&	0.0005\\
2&	1&	0&	2&	1&	1&	1&	-1&	0&	1&	0&	0&	18462.19332&	0.0005\\
2&	1&	0&	2&	1&	1&	1&	-1&	0&	1&	1&	1&	18497.13522&	0.0005\\
2&	-1&	0&	2&	2&	2&	1&	1&	0&	1&	1&	1&	22574.93438&	0.0005\\
2&	-1&	0&	2&	1&	1&	1&	1&	0&	1&	0&	0&	22691.9306&	0.0005\\
2&	-1&	0&	2&	1&	1&	1&	1&	0&	1&	1&	1&	22384.71706&	0.0005\\
2&	1&	1&	2&	2&	2&	1&	-1&	1&	1&	1&	1&	18280.36375&	0.0005\\
2&	1&	1&	2&	1&	1&	1&	-1&	1&	1&	0&	0&	18327.11545&	0.0005\\
2&	1&	1&	2&	1&	1&	1&	-1&	1&	1&	1&	1&	18364.30561&	0.0015\\
2&	-1&	1&	2&	1&	1&	1&	1&	1&	1&	1&	1&	22251.82489&	0.0005\\
2&	-1&	1&	2&	2&	2&	1&	1&	1&	1&	1&	1&	22442.64603&	0.0005\\
2&	-1&	1&	2&	1&	1&	1&	1&	1&	1&	0&	0&	22559.29702&	0.0005\\
2&	1&	2&	2&	2&	2&	1&	-1&	2&	1&	1&	1&	18146.86588&	0.0005\\
2&	1&	2&	2&	1&	1&	1&	-1&	2&	1&	0&	0&	18192.7778&	0.0005\\
2&	1&	2&	2&	1&	1&	1&	-1&	2&	1&	1&	1&	18232.1757&	0.0005\\
2&	-1&	2&	2&	1&	1&	1&	1&	2&	1&	1&	1&	22119.19787&	0.0005\\
2&	-1&	2&	2&	2&	2&	1&	1&	2&	1&	1&	1&	22310.60552&	0.0005\\
2&	-1&	2&	2&	1&	1&	1&	1&	2&	1&	0&	0&	22426.91413&	0.001\\
2&	1&	3&	2&	1&	1&	1&	-1&	3&	1&	0&	0&	18059.1839&	0.001\\
2&	1&	3&	2&	2&	2&	1&	-1&	3&	1&	1&	1&	18014.09752&	0.0005\\
2&	1&	3&	2&	1&	1&	1&	-1&	3&	1&	1&	1&	18100.74795&	0.0005\\
2&	-1&	3&	2&	1&	1&	1&	1&	3&	1&	1&	1&	21986.83747&	0.001\\
2&	-1&	3&	2&	2&	2&	1&	1&	3&	1&	1&	1&	22178.81699&	0.0005\\
2&	1&	4&	2&	2&	2&	1&	-1&	4&	1&	1&	1&	17882.06356&	0.0005\\
2&	1&	4&	2&	1&	1&	1&	-1&	4&	1&	0&	0&	17926.33965&	0.0005\\
2&	-1&	4&	2&	1&	1&	1&	1&	4&	1&	1&	1&	21854.74595&	0.0005\\
2&	-1&	4&	2&	2&	2&	1&	1&	4&	1&	1&	1&	22047.28122&	0.0007\\
2&	1&	5&	2&	2&	2&	1&	-1&	5&	1&	1&	1&	17750.76424&	0.0005\\
2&	1&	5&	2&	1&	1&	1&	-1&	5&	1&	0&	0&	17794.24209&	0.0005\\
2&	-1&	5&	2&	2&	2&	1&	1&	5&	1&	1&	1&	21916.00098&	0.0005\\
2&	1&	6&	2&	2&	2&	1&	-1&	6&	1&	1&	1&	17620.20372&	0.0005\\
2&	-1&	6&	2&	2&	2&	1&	1&	6&	1&	1&	1&	21784.97929&	0.0005\\
2&	1&	7&	2&	2&	2&	1&	-1&	7&	1&	1&	1&	17490.38923&	0.0005\\
1&	1&	10&	1&	0&	1&	1&	-1&	10&	1&	1&	1&	3187.48749&	0.002\\
1&	1&	10&	1&	0&	1&	1&	-1&	10&	1&	1&	2&	3219.81373&	0.00075\\
2&	-1&	10&	2&	2&	2&	2&	1&	10&	2&	2&	2&	4455.45403&	0.0025\\
2&	-1&	10&	2&	2&	3&	2&	1&	10&	2&	2&	3&	4699.22651&	0.0025\\
1&	1&	10&	1&	1&	1&	1&	-1&	10&	1&	0&	1&	8495.00218&	0.00075\\
3&	1&	10&	3&	3&	3&	3&	-1&	10&	3&	3&	3&	8620.54753&	0.001\\
1&	1&	10&	1&	1&	2&	1&	-1&	10&	1&	0&	1&	8687.20982&	0.00075\\
2&	-1&	10&	2&	1&	1&	2&	1&	10&	2&	1&	1&	11682.52107&	0.00075\\
2&	-1&	10&	2&	1&	1&	2&	1&	10&	2&	1&	2&	11715.3703&	0.00075\\
2&	-1&	10&	2&	1&	2&	2&	1&	10&	2&	1&	1&	11867.64154&	0.001\\
2&	-1&	10&	2&	1&	2&	2&	1&	10&	2&	1&	2&	11900.48704&	0.00075\\
2&	1&	10&	2&	1&	2&	1&	-1&	10&	1&	1&	1&	14430.18298&	0.00075\\
2&	1&	10&	2&	1&	2&	1&	-1&	10&	1&	1&	2&	14462.51043&	0.00075\\
2&	1&	10&	2&	1&	1&	1&	-1&	10&	1&	1&	1&	14463.03111&	0.00075\\
2&	1&	10&	2&	1&	1&	1&	-1&	10&	1&	1&	2&	14495.35797&	0.00075\\
3&	1&	10&	3&	2&	2&	3&	-1&	10&	3&	2&	2&	15865.18882&	0.00075\\
3&	1&	10&	3&	2&	3&	3&	-1&	10&	3&	2&	3&	16108.7271&	0.00075\\
2&	1&	10&	2&	2&	3&	1&	-1&	10&	1&	1&	2&	18333.50131&	0.00075\\
2&	1&	10&	2&	2&	2&	1&	-1&	10&	1&	1&	1&	18380.87112&	0.00075\\
2&	1&	10&	2&	2&	2&	1&	-1&	10&	1&	1&	2&	18413.19816&	0.00075\\
2&	-1&	10&	2&	2&	2&	1&	1&	10&	1&	1&	2&	22377.83416&	0.00075\\
2&	-1&	10&	2&	2&	3&	1&	1&	10&	1&	1&	2&	22541.91225&	0.00075\\
2&	-1&	10&	2&	2&	2&	1&	1&	10&	1&	1&	1&	22570.04272&	0.00075\\
2&	1&	10&	2&	1&	2&	1&	-1&	10&	1&	0&	1&	22658.90179&	0.00075\\
2&	1&	10&	2&	1&	1&	1&	-1&	10&	1&	0&	1&	22691.74864&	0.00075\\
2&	-1&	10&	2&	1&	1&	1&	1&	10&	1&	0&	1&	22958.0652&	0.00075\\
2&	-1&	10&	2&	1&	2&	1&	1&	10&	1&	0&	1&	23143.18458&	0.00075\\
2&	-1&	10&	2&	1&	1&	1&	1&	10&	1&	1&	2&	25687.06011&	0.00075\\
2&	-1&	10&	2&	1&	2&	1&	1&	10&	1&	1&	2&	25872.17892&	0.00075\\
2&	-1&	10&	2&	1&	1&	1&	1&	10&	1&	1&	1&	25879.2673&	0.00075\\
2&	-1&	10&	2&	1&	2&	1&	1&	10&	1&	1&	1&	26064.38734&	0.00075\\
2&	1&	11&	2&	1&	2&	1&	-1&	11&	1&	1&	2&	14329.36983&	0.00075\\
2&	1&	11&	2&	2&	3&	1&	-1&	11&	1&	1&	2&	18199.42792&	0.00075\\
2&	1&	11&	2&	2&	2&	1&	-1&	11&	1&	1&	1&	18246.41423&	0.00075\\
2&	-1&	11&	2&	2&	3&	1&	1&	11&	1&	1&	2&	22416.68639&	0.00075\\
2&	-1&	11&	2&	2&	2&	1&	1&	11&	1&	1&	1&	22444.02685&	0.00075\\
2&	1&	11&	2&	1&	2&	1&	-1&	11&	1&	0&	1&	22556.93031&	0.001\\
2&	-1&	11&	2&	1&	2&	1&	1&	11&	1&	0&	1&	23009.8026&	0.001\\
2&	1&	12&	2&	1&	2&	1&	-1&	12&	1&	1&	2&	14197.29144&	0.00075\\
2&	1&	12&	2&	2&	3&	1&	-1&	12&	1&	1&	2&	18066.08815&	0.00075\\
2&	1&	12&	2&	2&	2&	1&	-1&	12&	1&	1&	1&	18112.69634&	0.00075\\
2&	-1&	12&	2&	2&	3&	1&	1&	12&	1&	1&	2&	22291.81377&	0.00075\\
2&	-1&	12&	2&	2&	2&	1&	1&	12&	1&	1&	1&	22318.36426&	0.00075\\
2&	1&	13&	2&	2&	3&	1&	-1&	13&	1&	1&	2&	17933.48208&	0.002\\
2&	1&	13&	2&	2&	2&	1&	-1&	13&	1&	1&	1&	17979.72033&	0.00075\\
2&	-1&	13&	2&	2&	3&	1&	1&	13&	1&	1&	2&	22167.29671&	0.00075\\
2&	-1&	13&	2&	2&	2&	1&	1&	13&	1&	1&	1&	22193.05379&	0.00075\\
2&	1&	14&	2&	2&	3&	1&	-1&	14&	1&	1&	2&	17801.61828&	0.002\\
2&	1&	14&	2&	2&	2&	1&	-1&	14&	1&	1&	1&	17847.48916&	0.001\\
2&	1&	11&	2&	1&	1&	1&	-1&	11&	1&	0&	1&	22590.58422&	0.001\\
2&	1&	15&	2&	2&	3&	1&	-1&	15&	1&	1&	2&	17670.49357&	0.001\\
1&	1&	20&	1&	0&	0&	1&	-1&	20&	1&	1&	1&	3925.89105&	0.0015\\
1&	1&	20&	1&	1&	1&	1&	-1&	20&	1&	0&	0&	4198.15822&	0.0015\\
1&	1&	20&	1&	1&	1&	1&	-1&	20&	1&	1&	1&	4233.0993&	0.0007\\
2&	-1&	20&	2&	1&	1&	2&	1&	20&	2&	1&	1&	8124.06457&	0.001\\
2&	-1&	20&	2&	1&	1&	2&	1&	20&	2&	2&	2&	8206.61122&	0.003\\
2&	-1&	20&	2&	2&	2&	2&	1&	20&	2&	1&	1&	8314.28438&	0.0007\\
2&	-1&	20&	2&	2&	2&	2&	1&	20&	2&	2&	2&	8396.83017&	0.001\\
2&	1&	20&	2&	2&	2&	1&	-1&	20&	1&	1&	1&	18429.54268&	0.0005\\
2&	1&	20&	2&	1&	1&	1&	-1&	20&	1&	0&	0&	18477.14786&	0.0005\\
2&	1&	20&	2&	1&	1&	1&	-1&	20&	1&	1&	1&	18512.09022&	0.0005\\
2&	-1&	20&	2&	2&	2&	1&	1&	20&	1&	1&	1&	22593.27178&	0.0005\\
2&	-1&	20&	2&	1&	1&	1&	1&	20&	1&	0&	0&	22710.26738&	0.0005\\
2&	-1&	20&	2&	1&	1&	1&	1&	20&	1&	1&	1&	22403.054&	0.0005\\
2&	1&	21&	2&	2&	2&	1&	-1&	21&	1&	1&	1&	18295.15526&	0.0005\\
2&	1&	21&	2&	1&	1&	1&	-1&	21&	1&	0&	0&	18341.90629&	0.0005\\
2&	1&	21&	2&	1&	1&	1&	-1&	21&	1&	1&	1&	18379.09649&	0.0015\\
2&	-1&	21&	2&	1&	1&	1&	1&	21&	1&	1&	1&	22270.00064&	0.0005\\
2&	-1&	21&	2&	2&	2&	1&	1&	21&	1&	1&	1&	22460.82147&	0.0005\\
2&	-1&	21&	2&	1&	1&	1&	1&	21&	1&	0&	0&	22577.472&	0.0005\\
2&	1&	22&	2&	2&	2&	1&	-1&	22&	1&	1&	1&	18161.49498&	0.0005\\
2&	1&	22&	2&	1&	1&	1&	-1&	22&	1&	0&	0&	18207.40547&	0.001\\
2&	1&	22&	2&	1&	1&	1&	-1&	22&	1&	1&	1&	18246.8037&	0.001\\
2&	-1&	22&	2&	1&	1&	1&	1&	22&	1&	1&	1&	22137.21082&	0.001\\
2&	1&	23&	2&	2&	2&	1&	-1&	23&	1&	1&	1&	18028.56596&	0.0005\\
2&	1&	23&	2&	1&	1&	1&	-1&	23&	1&	0&	0&	18073.65005&	0.0005\\
2&	1&	24&	2&	2&	2&	1&	-1&	24&	1&	1&	1&	17896.3704&	0.0005\\
2&	1&	25&	2&	2&	2&	1&	-1&	25&	1&	1&	1&	17764.91321&	0.0005\\
2&	-1&	30&	2&	2&	2&	2&	1&	30&	2&	2&	2&	8403.7241&	0.0006\\
2&	1&	30&	2&	2&	2&	1&	-1&	30&	1&	1&	1&	18444.7873&	0.0005\\
2&	1&	30&	2&	1&	1&	1&	-1&	30&	1&	0&	0&	18492.39223&	0.0005\\
2&	1&	30&	2&	1&	1&	1&	-1&	30&	1&	1&	1&	18527.33549&	0.0005\\
2&	-1&	30&	2&	1&	1&	1&	1&	30&	1&	1&	1&	22421.74365&	0.0005\\
2&	-1&	30&	2&	2&	2&	1&	1&	30&	1&	1&	1&	22611.96289&	0.0005\\
2&	-1&	30&	2&	1&	1&	1&	1&	30&	1&	0&	0&	22728.95686&	0.0005\\
 \end{longtable}

\end{document}